\documentclass[a4paper, 12pt]{article}
\usepackage{amsmath}
\usepackage{amssymb}
\usepackage{epsfig}

\title{Four-dimensional SYM probes in wrapped M5-brane backgrounds}
\author{Jos\'e S\'anchez Loureda$^1$\footnote{email: j.m.sanchez-loureda@durham.ac.uk}
\quad and \quad
Douglas J. Smith$^2$\footnote{email: Douglas.Smith@durham.ac.uk} }

\begin{document}

\maketitle

\begin{center}

{\em Department of Mathematical Sciences,
University of Durham,
Science Laboratories,
South Rd,
Durham, DH1 3LE,
UK.}

\end{center}

\vspace{1.4cm}

\begin{abstract}

We study the worldvolume supersymmetric gauge theory of M-branes probing backgrounds corresponding to wrapped M5-branes. In the case of M5-branes wrapping a 2-cycle in $\mathbf{C}^{2}$, we use M2-brane probes to compute the BPS spectra of the corresponding $\mathcal{N} = 2$ gauge theory as well as M5-brane probes to calculate field theory parameters such as the gauge coupling, theta angle and complex scalar moduli space metric. This background describes a large class of Hanany-Witten type models when dimensionally reduced to Type IIA 10d string theory. We calculate the instanton action using a D0-brane probe in this limit. For the case of M5-branes wrapping a 2-cycle in $\mathbf{C}^{3}$, we firstly show an alternative method to derive this solution involving the projection conditions and certain spinor bilinear differential equations. We also consider M5-brane probes of this background, and analyse the corresponding $\mathcal{N} = 1$ MQCD gauge theory parameters. In general there were no supergravity corrections to field theory parameters when compared to previous flat-space field theory analysis.

\end{abstract}

\vspace{-20cm}
\begin{flushright}
DCPT-05/43 \\
\end{flushright}

\thispagestyle{empty}

\newpage

\setcounter{page}{1}

\section{Introduction}
\label{Introduction}

The main idea of this work is to verify that probing a certain supergravity background can correctly reproduce features of the four-dimensional $\mathcal{N} = 1,2$ supersymmetric gauge theory of which it is dual at the near-horizon limit. In particular, we wish to see if there are any corrections to the usual flat-space field theory analysis of $\mathcal{N} = 1,2$ supersymmetric gauge theories when we use the supergravity solution of localised brane intersections~\cite{Fayyazuddin:1999zu,Fayyazuddin:2000em,Brinne:2000fh,Brinne:2000nf} for our probe analysis. This approach works for a wide class of such theories. So one can obtain information about field theory from a geometric approach, in the spirit of the gauge theory - gravity correspondence.

Starting from the discovery of D-branes as stable extended objects in string theory~\cite{Polchinski:1995mt}, much work has been done examining their properties. Various dualities uncovered the M-theory origin of these branes~\cite{Polchinski:1995df,Witten:1995ex,Hull:1994ys,Townsend:1995kk}, in what was a great unifying leap forward. From the properties of these M-branes, almost all of the known spectrum of D-branes in lower dimensions had a nice geometrical origin. Soon, more elaborate constructions involving D-branes began\cite{Witten:1995im}, making use of the worldvolume fields living on them as sources for boundaries of other D-branes. In particular, supergravity solutions of orthogonal intersections of branes were examined~\cite{Tseytlin:1996hi,Gauntlett:1997cv}, where some of the branes were smeared over the others worldvolumes, as well as branes intersecting at angles~\cite{Ohta:1997fr}.  

Another important breakthrough was the construction of the type IIB Hanany-Witten brane models~\cite{Hanany:1996ie}, which were later recast in Type IIA language to deal with four-dimensional supersymmetric gauge theories~\cite{Witten:1997sc}. These provided an excellent geometrical description of the important analysis that had been done earlier on the exact low-energy effective action of $\mathcal{N} = 2$ gauge theories~\cite{Seiberg:1994rs}.

Another significant advance was the discovery of the duality now known as the gauge theory-gravity correspondence, or AdS/CFT as in the first examples\cite{Maldacena:1997re}. This was a concrete realisation of the older idea of the holographic principle~\cite{'tHooft:1993gx}. Originally inspired by the properties of D3-branes, it related Type IIB string theory on an Anti-de Sitter (AdS) spacetime to a conformal field theory (CFT), in that case, $\mathcal{N} = 4$ supersymmetric, large $N$, SU($N$) Yang-Mills theory in four dimensions. Many more examples were subsequently put forward relating different backgrounds to different supersymmetric Yang-Mills theories~\cite{Gubser:1998bc,Witten:1998qj}. In particular, many examples of branes wrapped on numerous manifolds and their dual Yang-Mills description were analysed.

The outline of this paper is as follows. In Section~\ref{SolutionM5}, we quickly review the $\mathcal{N} = 2$ supergravity solution of localised brane intersections we shall be using for the analysis in the next sections. Section~\ref{M5probesection} begins with a quick summary of the M5-brane worldvolume action followed by a probe calculation which yields the K\"{a}hler metric for the kinetic term of the complex moduli scalar fields. There is also a simple example of a parallel brane probe. Section~\ref{M2probesection} contains a calculation to determine the BPS spectra of the $\mathcal{N} = 2$ supersymmetric gauge theory. This is illustrated in two different ways: by a conventional M2-brane probe calculation, and by using the projection conditions and Killing spinors to calculate the central charge of the BPS probe M2-brane. The two are shown to be equivalent. Section~\ref{N2SYM} contains the calculation of the Yang-Mills coupling and theta angle of the $\mathcal{N} = 2$ Super Yang-Mills theory living on the D4-brane probe, once the M5-brane probe has been dimensionally reduced. This section also includes an instanton action calculation as a D0-brane probe of the 10d Type IIA background. Section~\ref{N1SYM} contains the analogous $\mathcal{N} = 1$ calculations. We also include a derivation of the previously known $\mathcal{N} = 1$ supergravity solution~\cite{Brinne:2000nf} using a method involving the projection conditions, certain differential forms and their differential equations built from the Killing spinor equation. Finally, the last section summarises the results obtained and discusses their implications.

\section{$\mathcal{N} = 2$ supergravity solution of fully localised M5-branes}
\label{SolutionM5}

\subsection{Review}

In this section we review the eleven-dimensional $\mathcal{N} = 2$ supergravity solution of fully localised M5-brane intersections~\cite{Fayyazuddin:1999zu,Fayyazuddin:2000em,Brinne:2000fh}. Viewed from an M-theory perspective, this corresponds to an M5-brane with worldvolume $\mathbf{R^{(1,3)}} \times \Sigma$, where $\Sigma \subset \mathbf{C}^{2}$ is a Riemann surface. This is a holomorphic embedding which preserves $\mathcal{N} = 2$ (in $d=4$) supersymmetry. This configuration is related, in the appropriate near-horizon limit, to $\mathcal{N} = 2$ supersymmetric gauge theories by the $AdS/CFT$ correspondence~\cite{Maldacena:1997re}. 

From Seiberg and Witten's work~\cite{Seiberg:1994rs} we know that the exact low energy effective action of a large class of $\mathcal{N} = 2$ four-dimensional gauge theories can be described by a family of Riemann surfaces $\Sigma \subset \mathbf{C}^2$. These Seiberg-Witten curves encode information about the gauge theory such as the exact mass of BPS states. 

These Seiberg-Witten curves were later found to have a geometric description, found by Witten~\cite{Witten:1997sc}, in terms of a Hanany-Witten construction~\cite{Hanany:1996ie}. This provided a much more intuitive geometric interpretation of the field theory results and has proved to be a useful method of describing large classes of supersymmetric gauge theories.

The Hanany-Witten setup in 10d Type IIA involves D4-branes with worldvolume directions $01236$ ending on NS5-branes extended in the 012345 directions. All the branes are located at $x^{8}=x^{9}=x^{10}=0$. The NS5-branes are separated in the $x^{6}$ direction and the D4-branes can be infinite in this direction, finite by stretching between two NS5-branes or semi-infinite by ending on only one of them from either side. We can then express four real dimensions in terms of two complex coordinates $v=x^{4} + ix^{5}$ and $s=x^{6} + i x^{7}$, where $x^{7}$ is the eleventh dimension (a circle of radius $R$). In eleven dimensions (as $R > 0$), the intersection between D4 and NS5-branes is smoothed out and becomes an M5-brane wrapping a 2-cycle in $\mathbf{C}^{3}$, with embedding $\mathbf{R}^{(1,3)} \times \Sigma$. The Riemann surface $\Sigma$ can be determined, up to moduli, using the known asymptotic form.

This Riemann surface $\Sigma$ is in fact the Seiberg-Witten curve for the gauge theory. The Seiberg-Witten differential also has an M-theory derivation~\cite{Fayyazuddin:1997by} (see~\cite{Elitzur:1997fh,Elitzur:1997hc} for a comprehensive review of these constructions). The BPS states correspond to minimal M2-branes whose boundary is on the M5-brane. The mass of the M2-brane gives the mass of the corresponding BPS-saturated state. 

\subsection{Reconstructing the supergravity solution}

One method of finding the supergravity solution of a particular brane configuration consists of using the projection conditions for preservation of supersymmetry to constrain the metric and four-form field strength. 

The integrability of the Killing spinor equation then allows one to relate geometries that preserve some degree of supersymmetry to those that in addition solve the equations of motion~\cite{Gauntlett:2002fz}. Typically, a supersymmetric solution which in addition solves the Bianchi identity and equations of motion for the background field strength will satisfy almost all the constraints from the Einstein equations. For the case of a geometry admitting a null Killing vector $K=e^{+}$, for example, one just needs to impose $E_{++}=0$ to obtain a full supersymmetric solution~\cite{Gauntlett:2002fz}, where $E_{\mu\nu}=0$ refers to the Einstein equations. 

More recently, the whole program relating to G-structures~\cite{Gauntlett:2002fz,Gauntlett:2003wb} has been carried out in full. This is a group theoretic approach which has proven useful in classifying supersymmetric solutions in various dimensions, depending on whether the minimal background Killing spinor gives either a time-like or null Killing vector. We shall not go into the details of this approach here, but we note that both supergravity solutions used in this paper can also be derived from the G-structures approach~\cite{Martelli:2003ki}. Quite similar to these is the new method we employ to derive the $\mathcal{N} = 1$ supergravity solution~\cite{Brinne:2000nf} in the first part of Section~\ref{N1SYM}. For the moment we consider the original approach.

\subsubsection{Solving the Killing spinor equations}

In the original paper~\cite{Fayyazuddin:1999zu}, after imposing the expected symmetries of the solution for the metric,  namely $\mathbf{R^{(1,3)}} \times \Sigma \times \mathbf{R^{(3)}}$, the supersymmetry projection conditions were used to solve the Killing spinor equations 

\begin{equation}
\tilde{D}_{I}\epsilon = 0,
\end{equation}
where

\begin{equation}
\tilde{D}_{I}\epsilon \equiv \nabla_{I}\epsilon + \frac{1}{288}\left[\Gamma^{\; JKLM}_{I}-8\delta_{I}^{J}\Gamma^{KLM}\right]F_{JKLM}\epsilon
\label{KillingSpinEq}
\end{equation}
and $F$ is the four-form field strength of 11d supergravity.

We use the notation in~\cite{Smith:2002wn}, with $\Gamma_{M}$ for the spacetime Dirac gamma-matrices and $\hat{\Gamma}_{m}$ for the tangent-space gamma-matrices. These are related by the vielbein $e^{m}_{M}$ such that

\begin{equation}
g_{MN}=e^{m}_{M}e^{n}_{N} \eta_{mn} \; , \; \Gamma_{M}=e^{n}_{M}\hat{\Gamma}_{m} \; , \; \left\{ \Gamma_{M} , \Gamma_{N} \right\} =2g_{MN} \; , \; \left\{ \hat{\Gamma}_{m} , \hat{\Gamma}_{N} \right\} =2\eta_{mn} .
\end{equation}
The number of supersymmetries preserved by a p-brane configuration is given by the number of spinors $\epsilon$ which satisfy the equation  

\begin{equation}
\hat{\Gamma}\epsilon = \epsilon
\end{equation}
where we have the definitions $\hat{\Gamma} =  \epsilon^{\alpha_{1}\ldots\alpha_{p}} \Gamma_{M_{1}\ldots M_{p}} \partial_{\alpha_{1}}X^{M_{1}}\ldots \partial_{\alpha_{p}}X^{M_{p}}$ and \\
$\Gamma_{M_{1}\ldots M_{p}} = \frac{1}{p!} \Gamma_{\left[ M_{1}\ldots M_{p}\right]}$. The $X^M$ is the embedding of the p-brane in the background geometry and the $\alpha_{i}$ denote the world-volume coordinates.

If we consider for definiteness our $\mathcal{N} = 2$ example, we know that the asymptotic form of the embedding (i.e.\ the Type IIA Hanany-Witten model) should be two sets of orthogonally intersecting M5-branes with worldvolume directions $012345$ and $012367$. This corresponds to the projection conditions

\begin{eqnarray}
\hat{\Gamma}_{012345} \epsilon &=& \epsilon \\
\hat{\Gamma}_{012367} \epsilon &=& \epsilon .
\end{eqnarray}

There are, however, more projections compatible with these that do not break any further supersymmetry, such as, for example,

\begin{equation}
\hat{\Gamma}_{4567}\epsilon = \hat{\Gamma}_{012389(10)} \epsilon = - \epsilon .
\end{equation}
If we now define complex co-ordinates 
\begin{eqnarray}
v&=&z^{1}=x^{4}+ix^{5} \\
s&=&z^{2}=x^6 +ix^7 ,
\end{eqnarray}
then we can concisely express the above relations as

\begin{equation}
\hat{\Gamma}_{0123a\bar{b}}\epsilon = i \delta_{a\bar{b}}\epsilon
\label{ProjM5}
\end{equation}
with $\delta_{a\bar{b}}$ as the tangent-space metric with $\delta_{1\bar{1}}=\frac{1}{2}$ in our conventions. This restriction on $\epsilon$ means that the solution will preserve $\frac{1}{4}$ of the supersymmetry (or equivalently, eight supercharges), which corresponds to $\mathcal{N}=2$ in four dimensions. We use conventions where $ds^2 = 2 g_{M\bar{N}} dz^M dz^{\bar{N}} = \delta_{a\bar{b}} e^{a}_{M} \left( \overline{e^{b}_{N}} \right) dz^M dz^{\bar{N}}$ for complex Hermitian metrics.

Solving the Killing spinor equations with these projection conditions and metric ans\"{a}tze $\mathbf{R^{(1,3)}} \times \Sigma \times \mathbf{R^{(3)}}$ then lead to the following metric and field strength~\cite{Fayyazuddin:1999zu,Fayyazuddin:2000em}:

\begin{equation}
ds^2 = H^{-1/3} {dx^2}_{(1,3)} + 2 H^{-1/3} g_{M \bar{N}} dz^{M}dz^{\bar{N}} + H^{2/3} {dx^2}_{(3)}
\label{FSspacetime}
\end{equation}

\begin{eqnarray}
F_{M\bar{N}\alpha\beta} & = & i \epsilon_{\alpha\beta\gamma}\partial_{\gamma}g_{M\bar{N}} \nonumber \\
F_{M89(10)} & = & - 4 i \partial_{M} g \nonumber \\
F_{\bar{N}89(10)} & = & 4 i \partial_{\bar{N}} g \nonumber \\
g & = & \left( g_{v\bar{v}} g_{s{\bar{s}}} - g_{s\bar{v}} g_{v\bar{s}} \right).
\label{Field Strength}
\end{eqnarray}
We denote with Greek letters $\alpha,\beta,\gamma$ the totally transverse directions $8,9,10$, and capital letters $M,N$ for the complex co-ordinates $v,s$.

The metric $g_{M\bar{N}}$ is constrained to be K\"{a}hler, with (square root) determinant $g$, and $H=4g$ from asymptotic conditions. This is similar to what we would expect from the harmonic function rules~\cite{Tseytlin:1996hi} of orthogonally intersecting branes, but with extra off-diagonal components. These components are what effectively describe the brane in the relative transverse directions. The equation of motion for $F$ with a magnetic source $J$ is

\begin{equation}
dF = J = J_{M\bar{N}} dz^{M} \wedge dz^{\bar{N}} \wedge dx^{8} \wedge dx^{9} \wedge dx^{10},
\end{equation}
where 

\begin{equation}
J_{M\bar{N}} = -4i \left( \pi l_{P} \right)^{3} \left( {\partial}_{M} f \right) \left( \overline{{\partial}_{N} f} \right) {\delta}^{2} (f) {\delta}^{3} (r) 
\label{M5source}
\end{equation}
since the M5-brane is wrapped on a Riemann surface $\Sigma$, defined by a holomorphic function $f(v,s)=0$ at $r=0$, where $r$ denotes the radial coordinate for the totally transverse space $\mathbf{R^{(3)}}$. This results, in terms of the K\"{a}hler potential for $g_{M\bar{N}}$, $K$, in the equation

\begin{equation}
8g(K) + {\partial}_{\gamma} {\partial}_{\gamma} K = -4\left( \pi l_{P} \right) ^{3}
\left\vert f \right\vert ^{2} {\delta}^{2} (f) {\delta}^{3} (r) 
\label{M5Kequation}
\end{equation}
which is related to the Monge-Amp\`{e}re equation.

\subsubsection{Taking the near-horizon limit}

Once the brane construction of a particular gauge theory is known, one can try to describe the supergravity dual of the field theory. In the same spirit as the $AdS/CFT$ correspondence, we identify the field theory parameters which should be kept fixed while taking a limit to decouple gravity and string modes. 

Since we are interested in describing the gravity dual, we only need to solve these equations in the near-horizon limit. In this limit we keep the gauge couplings and masses fixed while taking $l_{P} \rightarrow 0$. Looking at the example of a Hanany-Witten type IIA setup examined in~\cite{Fayyazuddin:1999zu}, we have, for example, magnetically charged states represented by D2-branes stretched between the D4-branes and NS5-branes. Classically, they would have a mass

$$
\begin{array}{ccccc}
m & =& \frac{|v|L}{g_{s} ({\alpha}')^{3/2}} & = & \frac{|w|}{g^2_{YM}},
\end{array}
$$

where $|v|$ is the coordinate distance between two D4-branes, and $L$ is the distance between two NS5-branes. Thus in the limit where we keep $w=v/{\alpha}'$ and the Yang-Mills coupling constant $g_{YM}$ fixed, while taking ${\alpha}' \rightarrow 0$, the field theory states have finite mass. 

Concretely, it was found that the relevant scalings of the supergravity variables in M-theory units, by defining $w,t$ and $y$ as follows, are:

$$
\begin{array}{ccccc}
w & = & \frac{v}{{l_{s}}^{2}} & = & \frac{vR}{{l_{P}}^{3}} \\
t^{2} & = & \frac{r}{g_{s} {l_{s}}^{3}} & = & \frac{r}{{l_{P}}^{3}} \\
y & = & \frac{s}{R}. & &   \\
\end{array}
$$

Our expectations from the $AdS/CFT$ duality suggests, for a conformal theory in a Hanany-Witten setup, a solution of the form of a warped product of $AdS_{5}$ with a compact six-dimensional manifold $M_{6}$. Requiring that the metric~(\ref{FSspacetime}) can be written in this form places several constraints on the components of the K\"{a}hler metric $g_{M\bar{N}}$ which are not obviously related to the equations of motion. However, they are compatible and a solution has been found~\cite{Brinne:2000fh}. 

\subsubsection{More general $\mathcal{N} = 2$ supersymmetric theories}

Looking at the example of a conformal theory with two NS5-branes separated by $\frac{1}{{g^{2}}_{YM}}$ in the $y$-plane intersected (for gauge group $SU(N)$) by $N$ infinite D4-branes, the holomorphic function $f(w,y)$ which describes this geometry, lifted to 11d, is

$$f= \left( y - \frac{1}{2{g^{2}}_{YM}} \right) \left( y + \frac{1}{2{g^{2}}_{YM}} \right) w^{N}. $$

This generalises for an arbitrary Riemann surface $\Sigma$ (an arbitrary holomorphic function $f(w,y)$). In this case, the supergravity solution can be determined from the K\"{a}hler potential $K$ which is given by~\cite{Brinne:2000fh}

\begin{eqnarray}
K & = & \frac{\pi N}{2t^{2}} \ln\left(\frac{\sqrt{t^{4} + \left\vert F \right\vert ^{4}} + t^{2}} {\sqrt{t^{4} + \left\vert F \right\vert ^{4}} - t^{2}} \right) + \frac{1}{2} \left\vert G \right\vert ^{2}  \nonumber \\
F^2 & = & f^{1/N},
\end{eqnarray}
where in general $N$ is defined as the degree of $f$ as a polynomial in $w$. To find explicit solutions we need to solve

\begin{equation}
\left({\partial}_{y}F^{2}\right) \left( {\partial}_{w} G \right) - \left({\partial}_{w}F^{2}\right) \left( {\partial}_{y} G \right) = 1
\label{FGconstraint}
\end{equation}
to find $G$. Whether this is easy or not depends on $f$.

Geometrically, the variables $(F^{2}, G)$ can be thought of as local co-ordinates transverse and parallel to the M5-brane. The above equation is simply the statement that the Jacobian of the holomorphic co-ordinate transformation from $(w,y)$ to $(F^{2}, G)$ is equal to one. It is also the necessary condition for the metric

\begin{equation}
g_{M \bar{N}} \equiv 2 \left( {\partial}_{M} F^2 \right) \left( \overline{ {\partial}_{N} F^2 } \right) g + 1/2 \left( {\partial}_{M} G \right) \left( \overline{ {\partial}_{N} G } \right)
\end{equation}
to have determinant $g$. The source equations~(\ref{M5source}), (\ref{M5Kequation}) reduce to the condition that $g$ is a harmonic function in the five-dimensional transverse space with radial co-ordinate

$$\tilde{r} \equiv \sqrt{t^4 + \left\vert F \right\vert ^4}$$
so that $g= \frac {\pi N}{8 {\tilde{r}}^3 }$. 

These new co-ordinates appear to be naturally suited to describe this M5-brane configuration.

\section{M5-brane probe}
\label{M5probesection}
\subsection{Preliminaries}

In this section we will study the action of an M5-brane probe. Probing our background, which is the supergravity dual of $\mathcal{N} = 2 $ gauge theories, we can obtain information about various field theory phenomena. In particular, we will calculate the metric of the complex scalars kinetic term in the effective Lagrangian.

We will consider the worldvolume action of the probe in the supergravity background manifold we have just described, which, up to warp factors, is  of the form $\mathbf{R}^{1,3} \times Q^{4} \times \mathbf{R}^{3}$, where $\mathbf{R}^{1,3}$ is the four-dimensional Minkowski spacetime, $Q^4$is a four-manifold of $SU(2)$ holonomy, i.e.\ a hyper-K\"{a}hler manifold, and $\mathbf{R}^{3}$ is a three-dimensional Euclidean manifold. The M5-brane which sources this background has worldvolume $\mathbf{R}^{1,3} \times \Sigma $, with $\Sigma$ being a 2-cycle embedded holomorphically in $Q^{4} \subset \mathbf{C}^2$ with complex coordinates $F^2, G$. It is located at a point in $\mathbf{R}^{3}$, the totally transverse directions. Explicitly, the metric reads

\begin{equation}
ds^2 = H^{-1/3} {dx^2}_{(1,3)} + 2 H^{-1/3} g_{M \bar{N}} dz^{M}dz^{\bar{N}} + H^{2/3} {dx^2}_{(3)}
\label{FSspacetimeFG}
\end{equation}
where the metric of the hyper-K\"{a}hler manifold is given by

\begin{equation}
g_{M \bar{N}} \equiv 2 \left( {\partial}_{M} F^2 \right) \left( \overline{ {\partial}_{M} F^2 } \right) g + 1/2 \left( {\partial}_{M} G \right) \left( \overline{ {\partial}_{N} G } \right)
\label{SpacetimeN2metric}
\end{equation}
and $g$ is given by $g= \frac {\pi N}{8 {\tilde{r}}^3 }$. The spacetime indices $M,N=F^2,G$ run over the hyper-K\"{a}hler part of the metric.

\subsection{The M5-brane worldvolume action}

The dynamics of the M5-brane probe are determined by its worldvolume action, the so called PST action~\cite{Pasti:1997gx} (see also~\cite{Aganagic:1997zq}). In the PST formalism the worldvolume fields are a self-dual three-form field strength $H=dB_{2}$ and an auxiliary scalar field $a$ (the PST scalar). The action is the sum of three terms:

\begin{equation}
S = \tau_{5} \int d^6\sigma \left[ {\mathcal{L}}_{DBI} + {\mathcal{L}}_{\mathcal{H}\tilde{\mathcal{H}}} + {\mathcal{L}}_{WZ} \right]
\label{M5Action}
\end{equation}
where the tension of the M5-brane is denoted by $\tau_{5}$. In the action~(\ref{M5Action}) the worldvolume field strength $H$ is combined with the pullback $P[C^{(3)}]$ of the background three-form potential $C^{(3)}$ to form the field $\mathcal{H}$:   

$$\mathcal{H} = H - e^{-\phi} P[C^{(3)}].$$ 
We can also define the field $\tilde{\mathcal{H}}$ as follows:

\begin{equation}
\tilde{\mathcal{H}}^{mn} = \frac{1}{3!\sqrt{-\mbox{det} G}}\frac{1}{\sqrt{-(\partial a)^2}}{\epsilon}^{mnlpqr}{\partial}_{l}a \mathcal{H}_{pqr}
\end{equation}
with $G$ being the induced metric on the M5-brane worldvolume (see Appendix~\ref{Conventions} for the conventions used throughout this paper). 

The explicit expressions for the three terms in the action are:

\begin{equation}
{\mathcal{L}}_{DBI} = -\sqrt{- \det \left( G_{mn} + i {\tilde{\mathcal{H}}}_{mn} \right)}
\end{equation}

\begin{equation}
{\mathcal{L}}_{\mathcal{H}\tilde{\mathcal{H}}} = \frac{1}{24(\partial a)^2}{\epsilon}^{lmnpqr} \mathcal{H}_{pqr} \mathcal{H}_{mns} G^{st} {\partial}_{l}a {\partial}_{t}a
\end{equation}

\begin{equation}
{\mathcal{L}}_{WZ} = \frac{1}{6!} {\epsilon}^{lmnpqr} \left[ P[C^{(6)}]_{lmnpqr} + 10 \mathcal{H}_{lmn} P[C^{(3)}]_{pqr} \right] .
\end{equation}

As discussed in~\cite{Pasti:1997gx}, the scalar field $a$ is an auxiliary field, which, by fixing its gauge symmetry, can be eliminated from the action at the expense of losing manifest covariance. We will not fix it for the time being.

\subsection{Probe calculation of complex scalar kinetic terms}
\label{ComScaKinTermsProbeN2}

Our M5-brane probe will have a worldvolume of the form $\mathbf{R}^{(1,3)} \times \Lambda$ where $\Lambda$ is a two-dimensional surface in $Q^4 \subset \mathbf{C}^2$ which is allowed to vary over $\mathbf{R}^{(1,3)}$. Also, we assign to it the worldvolume co-ordinates ${\sigma}^m$, $z$, $\bar{z}$ whose embeddings are holomorphic and of the form:

$$\begin{array}{rcl}
X^m & = & {\sigma}^{m}  \\
X^M & = & X^M \left( z, {\sigma}^{m} , u_{\alpha}\left({\sigma}^{m}\right)\right) \\
X^{\bar{N}} & = & X^{\bar{N}} \left( \bar{z}, {\sigma}^{m} , u_{\bar{\beta}}\left({\sigma}^{m}\right) \right) \\
X^{a} & = & X^{a} \left( z,\bar{z}, {\sigma}^{m} , u_{\alpha}\left({\sigma}^{m}\right) ,  u_{\bar{\beta}}\left({\sigma}^{m}\right) \right).
\end{array}$$

For the purposes of this calculation we define $m=0\ldots 3$, $M,N=F^2,G$ and $X^{a}$ refers to the totally transverse directions $a = 8,9,10$ (the conventions are similar to those in~\cite{deBoer:1997zy} where a related calculation was performed). Also, $z$, $\bar{z}$ are arbitrary co-ordinates on the Riemann surface $\Lambda$ which has $u_{\alpha}$, $u_{\bar{\beta}}$ as its complex moduli. 

We consider only small deviations from a supersymmetric embedding of the probe, so $\frac{\partial X^{M}}{\partial \sigma^{m}}$, $\frac{\partial X^{a}}{\partial z}$, $\frac{\partial X^{a}}{\partial \sigma^{m}}$, $\frac{\partial X^{a}}{\partial u_{\alpha}}$ and $\frac{\partial u_{\alpha}}{\partial \sigma^{m}}$ are small. This typically breaks all the supersymmetries, but since these are only very small deviations from the supersymmetric configuration we can expand the M5-brane probe action to quadratic order in these terms to find the metric on the moduli space.

As the five-brane action is invariant under world-volume diffeomorphisms, we can always choose $z$ and $\bar{z}$ in such a way that the induced metric on the Riemann surface is conformal, i.e.\ $g_{zz} = g_{\bar{z}\bar{z}} = 0$. As this will simplify things considerably, we will from now on assume this to be the case.

The first case we shall consider is a flat M5-brane probe with no worldvolume $H$ field turned on and neglecting the WZ contribution of the action. We will also ignore the $z,\bar{z}$ dependence of the $X^{a}$. In this case the probe action reduces to 

$$S = - \tau_{5} \int d^6\sigma \sqrt{-\mbox{det} \left( G_{6} \right)},$$
with $G_{6}$ the full 6d worldvolume metric. Explicitly, the action induced from the background metric becomes:

\begin{equation}
S = \int d^4\sigma d^2z 2 H^{-1} g_{z\bar{z}} \sqrt{-\mbox{det}\left({\eta}_{mn} + L_{mn}\right)}
\label{Worldvolume ActionDBI}
\end{equation}
where

\begin{eqnarray}
L_{mn} & = & 2 \left[ {\partial}_{m} X^M {\partial}_{n} X^{\bar{N}} \left( g_{M\bar{N}} -  g_{M\bar{z}}\frac{1}{g_{z\bar{z}}}g_{z\bar{N}} \right) + {\partial}_{m} u_{\alpha} {\partial}_{n} X^{\bar{N}} \left(  g_{\alpha \bar{N}} -  g_{\alpha \bar{z}}\frac{1}{g_{z\bar{z}}}g_{z\bar{N}} \right)  \right. \nonumber \\
&  & \left. {} + {\partial}_{m}X^M {\partial}_{n} u_{\bar{\beta}}  \left(  g_{M \bar{\beta}} -  g_{M \bar{z}} \frac{1}{g_{z\bar{z}}}g_{z\bar{\beta}} \right) +  {\partial}_{m} u_{\alpha} {\partial}_{n} u_{\bar{\beta}} \left(  g_{\alpha \bar{\beta}} -  g_{\alpha \bar{z}} \frac{1}{g_{z\bar{z}}}g_{z\bar{\beta}} \right)  \right] \nonumber \\
& {} & \; + \; g\left[ p_{\alpha \beta }\frac{\partial u_{\alpha}}{\partial {\sigma}^{m}} \frac{\partial u_{\beta}}{\partial {\sigma}^{n}} + p_{\alpha \bar{\beta}} \frac{\partial u_{\alpha}}{\partial {\sigma}^{m}} \frac{\partial u_{\bar{\beta}}}{\partial {\sigma}^{n}} + p_{\bar{\beta} \alpha }\frac{\partial u_{\bar{\beta}}}{\partial {\sigma}^{m}} \frac{\partial u_{\alpha}}{\partial {\sigma}^{n}} \right. \nonumber \\
& {} & \; + \; p_{\bar{\alpha} \bar{\beta}}\frac{\partial u_{\bar{\alpha}}}{\partial {\sigma}^{m}} \frac{\partial u_{\bar{\beta}}}{\partial {\sigma}^{n}} + p_{\alpha b }\frac{\partial u_{\alpha}}{\partial {\sigma}^{m}} \frac{\partial X^b}{\partial {\sigma}^{n}} + p_{\bar{\beta} b }\frac{\partial u_{\bar{\beta}}}{\partial {\sigma}^{m}} \frac{\partial X^b}{\partial {\sigma}^{n}} \nonumber \\
& {} & \left. {} + p_{a \alpha } \frac{\partial X^a}{\partial {\sigma}^{m}} \frac{\partial u_{\alpha}}{\partial{\sigma}^{n}}  + p_{a \bar{\beta} } \frac{\partial X^a}{\partial {\sigma}^{m}} \frac{\partial u_{\bar{\beta}}}{\partial {\sigma}^{n}}  + \delta_{ab} \frac{\partial X^a}{\partial {\sigma}^{m}} \frac{\partial X^b}{\partial {\sigma}^{n}} \right] 
\end{eqnarray}
and we have defined $g_{\alpha \bar{\beta}} \equiv \frac{\partial X^M}{\partial u_{\alpha}} g_{M\bar{N}} \frac{\partial X^{\bar{N}}}{\partial u_{\bar{\beta}}}$,
$ g_{M \bar{\beta}} \equiv g_{M\bar{N}} \frac{\partial X^{\bar{N}}}{\partial u_{\bar{\beta}}}$,
$p_{\alpha \bar{\beta}} \equiv \frac{\partial X^a}{\partial u_{\alpha}} {\delta}_{ab} \frac{\partial X^{b}}{\partial u_{\bar{\beta}}}$ and 
$p_{a \bar{\beta}} \equiv {\delta}_{ab} \frac{\partial X^{b}}{\partial u_{\bar{\beta}}}$.
In this notation, the spacetime metric $g_{M\bar{N}}$ is the same as that of Equation~(\ref{SpacetimeN2metric}).

The second bracket which is multiplied by $g$ only contributes terms which are at least cubic in small derivatives, so are typically higher order corrections and we will not analyse them here. The very last term is an exception since it remains quadratic but it nevertheless does not contribute to the effective four-dimensional theory we are interested in. 

The reason is that, on the one hand, if the complex space $\Lambda$ is non-compact, it would pick up an infinite mass term from the volume integral and could therefore be neglected in the four-dimensional field theory analysis. Whereas if the space $\Lambda$ is compact, we can in principle perform an expansion in terms of Fourier modes, and the fact that the endpoints of our probe are by definition constrained in the $X^{a}$ directions forces the zero modes to be at least linear in derivatives. These boundary conditions then imply that the last term will be of higher than quadratic order in derivatives, and therefore not contribute to our analysis. 

Upon expansion of the action~(\ref{Worldvolume ActionDBI}), the kinetic term for the scalars $X^{M}, X^{\bar{N}}$ and $u_{\alpha}$  reads

\begin{equation}
S_{kin} = \tau_{5} \int d^4\sigma d^2z \;  g^{-1} g_{z\bar{z}} \frac{1}{2} Tr(L).
\end{equation}
Looking at the quadratic terms in the complex moduli only, we find

\begin{equation}
S_{kin} = \tau_{5} \int d^4\sigma \; {\partial}_{m} u_{\alpha} {\partial}^{m} u_{\bar{\beta}} K_{\alpha \bar{\beta}},
\label{KinComScalars}
\end{equation}
where $K_{\alpha \bar{\beta}}$ is a K\"{a}hler metric given by

\begin{equation}
K_{\alpha \bar{\beta}} = {\int}_{\Lambda} d^{2}z \; g^{-1} \left( g_{\alpha \bar{\beta}} g_{z \bar{z}} -  g_{\alpha \bar{z}}  g_{z \bar{\beta}} \right).
\label{KmetricComKin}
\end{equation}
This is K\"{a}hler up to total derivative boundary terms of the form

$${\int}_{\Lambda} d^2z \left( \left( {\partial}_{\bar{z}} \bar{F}^2 \right) \left(  {\partial}_{\bar{\beta}} \bar{G} \right) - \left( {\partial}_{\bar{\beta}} \bar{F}^2 \right) \left( {\partial}_{\bar{z}} \bar{G} \right) \right) {\partial}_{z} \left[ \left( {\partial}_{\alpha} G \right) \left( {\partial}_{\gamma} F^2 \right) - \left( {\partial}_{\gamma} G \right) \left( {\partial}_{\alpha} F^2 \right) \right] $$
and

$${\int}_{\Lambda} d^2z \left( \left( {\partial}_{z} F^2 \right) \left( {\partial}_{\alpha} G \right) - \left( {\partial}_{\alpha} F^2 \right) \left( {\partial}_{z} G \right) \right) {\partial}_{\bar{z}} \left[ \left( {\partial}_{\bar{\gamma}} \bar{F}^2 \right) \left( {\partial}_{\bar{\beta}} \bar{G} \right) - \left( {\partial}_{\bar{\beta}} \bar{F}^2 \right) \left( {\partial}_{\bar{\gamma}} \bar{G} \right) \right] ,$$     
where we have ignored contributions coming from the $g ({\sum}_{x,y} p_{x y })$ terms (the totally transverse fluctuations of the brane), as explained above. In this calculation we have also explicitly used the fact that the spacetime metric $g_{M\bar{N}}$ is K\"{a}hler.

These terms can be written as a total derivative straight away since they are a product of holomorphic and anti-holomorphic factors. In particular, if we define the one-forms

\begin{equation}
\Phi_{\alpha} = \left[ \left( {\partial}_{z} F^2 \right) \left( {\partial}_{\alpha} G \right) - \left( {\partial}_{\alpha} F^2 \right) \left( {\partial}_{z} G \right) \right]dz ,
\end{equation}
then $d\Phi_{\alpha} = 0$ because of holomorphicity. So if we also define the scalars

\begin{equation}
B_{\alpha \gamma} = \left( {\partial}_{\alpha} G \right) \left( {\partial}_{\gamma} F^2 \right) - \left( {\partial}_{\gamma} G \right) \left( {\partial}_{\alpha} F^2 \right) ,
\end{equation}
the boundary terms are of the form

\begin{equation}
{\int}_{\Lambda} \; dB_{\alpha\gamma} \wedge \bar{\Phi}_{\bar{\beta}}  \; \mbox{and} \; {\int}_{\Lambda}  \; d\bar{B}_{\bar{\beta}\bar{\gamma}} \wedge \Phi_{\alpha}  .
\end{equation}
Evaluating them at the boundary results in

\begin{equation}
{\int}_{\partial \Lambda} \; B_{\alpha\gamma} \; \bar{\Phi}_{\bar{\beta}}  \; \; \mbox{and} \; \; {\int}_{\partial \Lambda} \; \bar{B}_{\bar{\beta}\bar{\gamma}} \; \Phi_{\alpha} .
\end{equation}

So we can impose that these terms vanish at the boundary. For a non-compact probe, the asymptotic embedding is independent of the moduli, so clearly these boundary terms vanish (since $\left( {\partial}_{\alpha} F^2 \right)=\left( {\partial}_{\alpha} G \right) = 0$ asymptotically), and hence the metric is K\"{a}hler. However, while a finite D4-brane probe in Type IIA can end on a background NS5-brane, this is only possible for a supersymmetric probe as an approximation for small $R > 0$, so for such a probe one should choose appropriate boundary conditions such that the boundary terms vanish.

Additionally, there are the mixed terms ${\partial}_{\mu} u_{\alpha} {\partial}^{\mu} X^{\bar{N}}$ and ${\partial}_{\mu}X^{M} {\partial}^{\mu} u_{\bar{\beta}}$ as well as the quadratic term of the complex scalars ${\partial}_{\mu}X^{M} 
{\partial}^{\mu}  X^{\bar{N}}$. They all have K\"{a}hler metrics on their moduli space with boundary terms similar in form to the ones we have just analysed, giving the expected result that the moduli space of all the complex scalars is given by a K\"{a}hler metric.

%
%
%

From the expression for the K\"{a}hler metric~(\ref{KmetricComKin}) of the complex scalars with respect to the complex moduli, one can then obtain the standard form of the scalar kinetic terms of the $\mathcal{N} = 2 $ effective Lagrangian in the usual ways (see for example~\cite{deBoer:1997zy}).

\subsection{A simple example: the parallel brane probe}

Another example is to probe the background with an M5-brane which is parallel to the background M5-brane configuration. This does not imply it is flat, but merely that it somehow reflects the shape of the background. We shall let our probe have worldvolume $0123z\bar{z}$, where $z=\sigma^4 + i \sigma^5$. This time we do not consider fluctuations in the complex moduli of the brane. We will let the probe have a time dependence on $Q^4 \times \mathbf{R}^3$. The embeddings are then 

$$
\begin{array}{rcl}
X^m & = & \sigma^m \\
X^M & = & X^M (z, \sigma^{0}) \\
X^{\bar{N}} & = & X^{\bar{N}} (\bar{z}, \sigma^{0}) \\
X^{a} & = & X^{a} (\sigma^{0}) .
\end{array} 
$$
The action for the kinetic scalar terms then becomes

\begin{equation}
S_{kin} = \tau_{5} \int d^{4}\sigma d^{2}z g^{-1} g_{z\bar{z}} \left( g_{0\bar{0}} - g_{0\bar{z}} \frac{1}{g_{z\bar{z}}} g_{z\bar{0}} + g \left( {\partial}_{0} X^{a} \right)^{2} \right)
\end{equation}
where $g_{0\bar{0}} = {\partial}_{0} X^{M} g_{M\bar{N}} {\partial}_{0} X^{\bar{N}}$.

Now, for a probe which is parallel to the background we set $z=G$ which simplifies the above expression to 

\begin{equation}
S_{kin} = \tau_{5} \int d^{4}\sigma d^{2}z \left( \left\vert {\partial}_{0} F^{2} \right\vert ^{2} + \frac{1}{2} \left( {\partial}_{0} X^{a} \right) ^{2} \right) .
\end{equation}

This means that the brane sees a flat metric on the transverse directions which agrees with the expectation of a flat moduli space metric. Also, there is a trivial volume form which seems to suggest that these co-ordinates are a natural way to describe this configuration.

We shall return to more results from M5-brane probes shortly, but before we do that, a quick foray into an M2-brane probing the BPS spectra of the field theory.

\section{M2-brane probe}
\label{M2probesection}
\subsection{Introductory remarks}

The main result of this section is to calculate the mass of BPS states in four-dimensional $\mathcal{N} = 2$ supersymmetric gauge theories. We shall be using an M2-brane as a probe of the supergravity background corresponding to completely localised M5-brane configurations in M-theory (or equivalently M5-branes wrapping 2-cycles in $\mathbf{C}^{2}$), which is the supergravity dual of a large class of such gauge theories. States corresponding to BPS monopoles are realized as two-branes ending on the five-branes. One example of this are the membranes in a Hanany-Witten type setup. In particular we check whether this method provides corrections to the previous flat-space four-dimensional $\mathcal{N} = 2$ supersymmetric field theory analysis~\cite{Henningson:1997hy,Mikhailov:1997jv}.

There are two ways in which this can be done. In the probe analysis, we find a suitable complex structure in the hyper-K\"{a}hler part of the background in which to embed the M2-brane holomorphically, and then proceed to calculate the induced volume. We calculate the case of a static M2-brane and check it receives no corrections from the supergravity description.

The other method is based on the approach of calibrations~\cite{Harvey:1982xk,Gibbons:1998hm,Gauntlett:1998vk,Acharya:1998en}. This relates the BPS bound to the central charge of the 11d supergravity supersymmetry algebra. We take into account the generalisation of these calibration forms to include arbitrary background fields~\cite{Hackett-Jones:2003vz}. Again these topological charges give no corrections to the previous flat-space field theory calculations of the BPS monopole mass.

In the following, we shall establish the complex structure the M2-brane probe should be embedded holomorphically with respect to, and proceed to calculate its worldvolume. There follows a brief review of the concept of generalised calibrations and a calculation of the calibration bound for the M2-brane given the appropriate supersymmetric projection conditions. In both cases, we find no corrections to the previous flat-space field theory analysis.

\subsection{M2-brane probe calculation}

In this section, we will study the action of an M2 brane probe since it is known that minimal area membranes which end on M5 branes are related to the BPS states of $\mathcal{N} = 2 $ gauge theories to which our supergravity background is dual. Our background is sourced by the M-theory configuration described in the last section, which has the topology $\mathbf{R}^{1,3} \times Q^{4} \times \mathbf{R}^{3} $ up to warp factors, where $ Q^{4}$ is a hyper-K\"{a}hler manifold.

\subsubsection{Preliminaries}

Consider an M2-brane probe with worldvolume $\mathbf{R} \times D $, where $D$, the spatial part of the M2-brane, is a two dimensional surface embedded in the manifold $Q^{4}$ given by our background. Apart from the warp factor, we know $Q^{4}$ is hyper-K\"{a}hler because all two-complex dimensional K\"{a}hler manifolds are automatically hyper-K\"{a}hler. This means that instead of the usual one complex structure, this geometry admits a family of inequivalent complex structures parametrised by a two-sphere $S^2$, with $SU(2)$ commutation relations between them. Also, in four dimensions, the hyper-K\"{a}hler condition implies Ricci flatness and should therefore admit a covariantly constant holomorphic two-form.

We denote by $\Sigma$ the surface of the M5-brane which is embedded holomorphically in $Q$. Now, we wish to embed our M2 probe holomorphically so that its spatial part has a boundary $C=\partial D$ that lies on $\Sigma$, i.e.\ so the two-brane ends on the  five-brane. To achieve this the M2-brane shall be embedded holomorphically with respect to some complex structure $J^\prime$ which is orthogonal to the complex structure $J$ in which the M5-brane was embedded holomorphically. Given a complex structure $J$, the set of such $J^\prime$ for a hyper-K\"{a}hler manifold is parametrised by an $S^1$ that actually corresponds to the phase of the central charge of the BPS saturated state~\cite{Henningson:1997hy}.

To further and completely distinguish between the different possibilities, we also require that the M2-brane probe satisfy the supersymmetry projection conditions. 

\subsubsection{Choosing the appropriate complex structure}
\label{ComplexStructureM2}

For our particular background geometry, the five-brane is wrapped around the holomorphic curve $\Sigma$ and the Killing spinors satisfy~\cite{Becker:1995kb}:

\begin{equation}
\hat{\Gamma}_{0123}\hat{\Gamma}_{a\bar{b}}\epsilon = i \delta_{a\bar{b}} \epsilon.
\end{equation}
with $a,b$ running over $1,2$. These projection conditions preserve 8 real components of $\epsilon$ and thus give $\mathcal{N}=2$ supersymmetry in four dimensions.

Introducing the two-brane which ends on the five-brane requires the additional constraint~\cite{Fayyazuddin:1997by}

\begin{equation}
\frac{1}{2}\epsilon^{\alpha\beta}\Gamma_{0}\Gamma_{I\bar{J}} \partial_{\alpha} W^{I} \partial_{\beta}W^{\bar{J}}\epsilon = \epsilon
\label{M2projCond}
\end{equation}
where $W^{I}$ now denotes the embedding of the two-brane with respect to a different complex structure.

Explicitly, if we rewrite the hyper-K\"{a}hler part of the metric in terms of the vielbeins

$$
g_{M\bar{N}}dz^{M}dz^{\bar{N}} = \vert dZ^{1} \vert^2 + \vert dZ^2 \vert^2 = e^{a}_{M}\left(\overline{{e}^{b}_{N}}\right)\delta_{a\bar{b}} dz^{M}dz^{\bar{N}},
$$
the complex structure $J$, compatible with the M5-brane configuration, becomes

\begin{eqnarray}
dZ^1 & = & Re \left( e^{1}_{M} dz^{M} \right) + i Im \left(  e^{1}_{M} dz^{M} \right) \nonumber \\
dZ^2 & = & Re \left( e^{2}_{M} dz^{M} \right) + i Im \left(  e^{2}_{M} dz^{M} \right) .
\end{eqnarray}
We can now deduce the alternative complex structure $J^\prime$  that satisfies the projection conditions and the orthogonality constraint. In terms of the differentials, these are

\begin{eqnarray}
dW^1 & = & Re \left( e^{1}_{M} dz^{M} \right) + i Re \left( e^{2}_{M} dz^{M} \right) \nonumber \\
dW^2 & = & Im \left(  e^{1}_{M} dz^{M} \right) - i Im \left(  e^{2}_{M} dz^{M} \right) .
\label{Wdiff}
\end{eqnarray}

The M2-brane probe shall be embedded holomorphically with respect to the co-ordinates $W^1,W^2$ in the above basis. As we will see, we won't actually need to integrate the $dW^1,dW^2$  differentials, which simplifies the task considerably. Additionally, one can also trivially include an arbitrary phase which rotates the $W^1,W^2$ co-ordinates. We include this phase for completeness in the analysis of Section~\ref{Section4}.

We can rewrite the M2-brane projection condition~(\ref{M2projCond}) in terms of the M5-brane holomorphic variables using Equation~(\ref{Wdiff}). In this language, the projection condition is

\begin{equation}
\left(\hat{\Gamma}_{0ab} + \hat{\Gamma}_{0\bar{a}\bar{b}}\right)\epsilon = \epsilon
\label{ProjM2nophase}
\end{equation}
with again $a,b=1,2$. This additional constraint cuts the number of supersymmetries by half (leaving four real supersymmetries), expressing the fact that the M2-brane is a BPS state in the worldvolume theory of the M5-brane.

\subsubsection{Probe calculation}

We shall now consider our background spacetime $\mathbf{R}^{1,3} \times Q^{4} \times \mathbf{R}^{3} $ with metric

\begin{equation}
ds^2 = H^{-1/3} {dx^2}_{(1,3)} + 2 H^{-1/3} g_{M \bar{N}} dz^{M}dz^{\bar{N}} + H^{2/3} {dx^2}_{(3)} ,
\label{M2metricN2}
\end{equation}
where

\begin{equation}
g_{M \bar{N}} \equiv 2 \left( {\partial}_{M} F^2 \right) \left( \overline{ {\partial}_{N} F^2 } \right) g + 1/2 \left( {\partial}_{M} G \right) \left( \overline{ {\partial}_{N} G } \right)
\end{equation}
and $g$ is given by $g= \frac {\pi N}{8 {\tilde{r}}^3 }$ . 

We define the spacetime indices  $m=0,\ldots ,3 $ and $M,N=F^2,G$ that run over the Lorentzian part and the hyper-K\"{a}hler part respectively. 

The worldvolume co-ordinates of our M2-brane shall be $( t, \sigma, \bar{\sigma} )$, where we have complexified the spatial part of the brane (with $\sigma = \sigma^{1} + i \sigma^{2}$) for future convenience. These will have holomorphic embeddings of the form

\begin{eqnarray}
X^0 & = & t \nonumber \\
W^I & = & W^I (\sigma) \nonumber \\
W^{\bar{J}} & = & W^{\bar{J}} (\bar{\sigma}) .
\end{eqnarray}

This static probe will provide information about the mass of BPS states of the dual gauge theory. The action of an M2-brane is given by

\begin{equation}
S_{M2} = - \tau_{2} \int d^{3}\sigma \sqrt{-\mbox{det}\left(G_{ij}\right)} + \int \Xi
\end{equation}
where $\tau_{2}$ is the tension, $G_{ij}$ is the pullback of the spacetime metric onto the two-brane and $\Xi$ is the pullback of the spacetime three-form potential. Note that this last term vanishes for our particular embedding and so does not contribute in the analysis.

In terms of real co-ordinates, we can define $z^{M} = x^{M} + i y^{M}$ and split the complex vielbein into real and imaginary parts $e^{a}_{M} = {\alpha}^{a}_{M} + i {\beta}^{a}_{M}$. The holomorphy condition on the worldvolume induced co-ordinates, 

$$\frac{\partial W^1}{\partial \bar{\sigma}} = \frac{\partial W^2}{\partial \bar{\sigma}} = 0$$ 
then gives a set of four constraints on the vielbeins. We can simplify these equations by defining

$$
\begin{array}{ccc}
A^{a}_{i} \equiv {\alpha}^{a}_{M} \partial_{i} x^{M} - {\beta}^{a}_{M}\partial_{i} y^{M} \\
B^{a}_{i} \equiv {\beta}^{a}_{M} \partial_{i} x^{M} + {\alpha}^{a}_{M}\partial_{i} y^{M}.
\end{array}
$$
In terms of these new variables, our holomorphy constraints imply 
$$
A^{1}_{1} = A^{2}_{2}, \; A^{1}_{2} = -A^{2}_{1} , \; B^{1}_{1} = -B^{2}_{2} , \; B^{1}_{2} = B^{2}_{1}.
$$
The induced metric can then be written
$$
G_{ij} = \delta_{a\bar{b}} \left[ A^{a}_{i} + i B^{a}_{i} \right] \left[ A^{b}_{j} - i B^{b}_{j} \right] + \left[ i \leftrightarrow j \right] .
$$
We can simplify this further by defining new complex variables
$$
C^{a}_{i} = A^{a}_{i} + i B^{a}_{i}
$$
which transforms the constraints to 

\begin{equation}
C^{1}_{1}=\overline{C^{2}_{2}} \;\; \mbox{and} \;\; C^{1}_{2}=-\overline{C^{2}_{1}}. 
\end{equation}
More concretely, in terms of our complex vielbeins, we have
\begin{equation}
C^{a}_{i} = e^{a}_{M} \frac{\partial z^M}{\partial \sigma^{i}}.
\end{equation}
Finally, the induced metric can be written in the form
\begin{equation}
G_{ij} = \delta_{a\bar{b}} e^{a}_{M} {e^{\bar{b}}_{\bar{N}} \frac{\partial z^M}{\partial \sigma^{i}}  \frac{\partial z^{\bar{N}}}{\partial \sigma^{j}}} + \left[ i \leftrightarrow j \right] = \delta_{a\bar{b}} C^{a}_{i} \left(\overline{C^{b}_{j}}\right) + \left[ i \leftrightarrow j \right]
\label{IndM2metric}
\end{equation}
which can be checked explicitly to be hermitian. In particular, one can evaluate the components of the induced metric. Equation~(\ref{IndM2metric}) reveals that $G_{12}=G_{21}=0$ and $G_{11}=G_{22}$. The precise form of the non-trivial components is

\begin{eqnarray}
G_{11} &=& 2g\vert \partial_{1} F^2 \vert^2 + \frac{1}{2} \vert \partial_{1} G \vert^2
\end{eqnarray}
with the notation $\partial_{1} \equiv \frac{\partial}{\partial \sigma^{1}}$. 

In terms of complex vielbein components, the holomorphy conditions of $W^i (\sigma)$ reduce to the following equations

\begin{eqnarray}
Re \left( e^{1}_{M} \frac{\partial z^{M}}{\partial \sigma^1}  \right) & = & Re \left( e^{2}_{M} \frac{\partial z^{M}}{\partial \sigma^2} \right) \nonumber \\
 Re \left( e^{1}_{M} \frac{\partial z^{M}}{\partial \sigma^2}  \right) & = & - Re \left( e^{2}_{M} \frac{\partial z^{M}}{\partial \sigma^1}  \right) \nonumber \\
Im \left( e^{1}_{M} \frac{\partial z^{M}}{\partial \sigma^1}  \right) & = & - Im \left( e^{2}_{M} \frac{\partial z^{M}}{\partial \sigma^2} \right) \nonumber \\
 Im \left( e^{1}_{M} \frac{\partial z^{M}}{\partial \sigma^2}  \right) & = & Im \left( e^{2}_{M} \frac{\partial z^{M}}{\partial \sigma^1}  \right) .
\end{eqnarray}
We can choose $e^{1}_{M} = 2 \sqrt{g} \partial_{M} F^2$ and $e^{2}_{M} = \partial_{M} G$ which leads to the constraints
\begin{eqnarray}
2 \sqrt{g} \frac{\partial F^2}{\partial \sigma^1} &=& \frac{\partial \bar{G}}{\partial \sigma^2} \nonumber \\
2 \sqrt{g} \frac{\partial F^2}{\partial \sigma^2} &=& - \frac{\partial \bar{G}}{\partial \sigma^1}
\label{HolRestraintsFG}
\end{eqnarray}
which are identical to our earlier results. If we now include the warp factor $2H^{-1/3}$ we had been ignoring until now and look at the full determinant of the induced metric on the M2-brane probe we conclude

\begin{eqnarray}
\int \sqrt{-\mbox{det}\left(G_{\mu\nu}\right)} \; dt\wedge d\sigma^1 \wedge d\sigma^2 &=& \int \sqrt{-2H^{-2/3}G_{00}G_{11}G_{22}} \; dt\wedge d\sigma^1 \wedge d\sigma^2 \nonumber \\
&=& \int \left( \frac{\partial F^2}{\partial \sigma^1}\frac{\partial G}{\partial \sigma^2} - \frac{\partial G}{\partial \sigma^1}\frac{\partial F^2}{\partial \sigma^2}\right) \; dt\wedge d\sigma^1 \wedge d\sigma^2 \nonumber \\
& = & \int dt\wedge dF^2 \wedge dG.
\end{eqnarray}
If we consider the spatial part of the probe, this induced worldvolume integral times the tension of the brane results in a probe mass given by

\begin{equation}
\mbox{Mass} = \left\vert \int dF^2 \wedge dG \right\vert .
\label{MassofM2probe}
\end{equation}

This gives a very natural frame in which to describe the M2-brane probe dynamics. In some sense, we have chosen the appropriate co-ordinates so the induced probe brane action has a trivial (in $g$) volume form. This is similar to what happened in the previous M5-brane example of the complex scalar kinetic terms~(\ref{KinComScalars}).

\subsubsection{Check from topological arguments}

On a different note, one can check that the result is correct by calculating the induced K\"{a}hler form $K_{D}$ and holomorphic two form $\Omega_{D}$ on the spatial part of the M2-brane probe. We follow closely the methods of~\cite{Henningson:1997hy} (see also~\cite{Mikhailov:1997jv}) which analysed the case of M2-brane and M5-brane intersections in flat space, without taking into account the full M5-brane background geometry.  

Our previous results should agree with those deduced from a topological perspective. To preserve the required amount of supersymmetry, we must require the spatial volume element of the surface $D$ to be minimised so that it saturates a topological bound. Given our choice of complex structure $J^\prime$ on $Q^{4}$, a useful identity is~\cite{Henningson:1997hy}

\begin{equation}
\frac{1}{4} \left(\left(\ast K_{D} \right)^2 + \left\vert \ast \Omega_{D} \right\vert ^2 \right) = 1
\end{equation}
where $K_{D}$ is the pullback of the K\"{a}hler form $K$ to $D$, and the $\ast$ denotes the Hodge dual with respect to the induced metric on $D$. 

The area $A_{D}$ of the spatial part of the two-brane fulfils the inequalities

\begin{equation} 
2 A_{D} = 2 \int_{D} V_{D} = \int_{D} V_{D} \sqrt{\left(\ast K_{D} \right)^2 + \left\vert \ast \Omega_{D} \right\vert ^2 } \geq \int_{D} \left\vert \ast \Omega_{D} \right\vert V_{D} \geq \left\vert \int_{D} \Omega_{D} \right\vert
\end{equation}
where $V_{D}$ is the volume-form of $D$. The first inequality is saturated if and only if the pull-back $K_{D}$ of the K\"{a}hler form vanishes, while the second condition requires that the phase of the pullback $\ast\Omega_{D}$ is constant over $D$. The surface $D$ is then a holomorphic embedding with respect to some complex structure $J^\prime$ which is orthogonal to the complex structure $J$. 

Explicitly, the pullback of the two-form $\Omega_{D}$ is

\begin{eqnarray}
\Omega_{D} &=& 2H^{-1/3} P \left( e_{F} \wedge e_{G} \right) \nonumber \\
&=& 2^{1/3}g^{1/6} \left( \frac{\partial F^2}{\partial \sigma^1}\frac{\partial G}{\partial \sigma^2} - \frac{\partial G}{\partial \sigma^1}\frac{\partial F^2}{\partial \sigma^2}\right) d\sigma^1 \wedge d\sigma^2 .
\end{eqnarray}
The area of the spatial part $D$ of the probe is given by

\begin{eqnarray}
\label{AreaofSpatialM2}
\int \sqrt{\mbox{det}\left( {G_{M2}}^D \right)} \; d\sigma^1 \wedge d\sigma^2 &=& \int \sqrt{G_{11}G_{22}} \; d\sigma^1 \wedge d\sigma^2   \\
&=& 2^{1/3} \int g^{1/6} \left( \frac{\partial F^2}{\partial \sigma^1}\frac{\partial G}{\partial \sigma^2} - \frac{\partial G}{\partial \sigma^1}\frac{\partial F^2}{\partial \sigma^2}\right) d\sigma^1 \wedge d\sigma^2 \nonumber 
\end{eqnarray}
so we have 

\begin{equation}
\int \Omega_{D} \wedge \bar\Omega_{D} = \left( V_{D} \right)^2 = \int \det \left( {G_{M2}}^{D} \right) .
\end{equation}
We can quickly check that the induced K\"{a}hler form $K_{D}$ on the spatial part of the M2-brane probe

\begin{eqnarray}
K_{D} &=& 2H^{-1/3} \left( 2g \; dF^2 \wedge d\bar{F}^2 + 1/2 \; dG \wedge d\bar{G} \right) \nonumber \\
&=& 0
\end{eqnarray}
vanishes identically. This follows from the holomorphy constraints~(\ref{HolRestraintsFG}). It provides a check on the equivalence of the holomorphic two-form and the volume element of the probe as deduced from topological arguments.

Now, in order to compare Equations~(\ref{AreaofSpatialM2}) and~(\ref{MassofM2probe}), we have to realise that the energy-momentum is an invariant quantity. The area element of the probe we have just calculated is not an invariant quantity since it doesn't include the time component. So we need to add a factor of $\sqrt{-g_{00}}=H^{-1/6}=2^{-1/3}g^{-1/6}$ to Equation~(\ref{AreaofSpatialM2}). Doing this then gives the invariant mass term

\begin{eqnarray}
\mbox{Mass} &=&  2^{1/3} \int \sqrt{-g_{00}} g^{1/6} \left( \frac{\partial F^2}{\partial \sigma^1}\frac{\partial G}{\partial \sigma^2} - \frac{\partial G}{\partial \sigma^1}\frac{\partial F^2}{\partial \sigma^2}\right) d\sigma^1 \wedge d\sigma^2 \nonumber \\
&=&  \left\vert \int dF^2 \wedge dG \right\vert 
\end{eqnarray}
which then agrees with Equation~(\ref{MassofM2probe}). 

This also equivalent to the recent result in~\cite{Fayyazuddin:2005pm} which used a slightly different method and notation. The M2-brane probe satisfies the same calibration bound in both cases. The next section is a spinorial derivation of this bound from the supersymmetry projection conditions. 

\subsection{Spinorial derivation of the M2-brane BPS bound}
\label{Section4}
\subsubsection{General form of supersymmetry algebra for membranes}

Recently, the understanding of the general structure of supersymmetric solutions of supergravity theories has made great strides (see for example~\cite{Hackett-Jones:2003vz} and references therein). This stems from careful analysis of the Killing spinor equations:

\begin{equation}
\tilde{D}_{M}\epsilon = 0
\end{equation}
where

\begin{equation}
\tilde{D}_{M}\epsilon \equiv \nabla_{M}\epsilon + \frac{1}{288}\left[\Gamma^{\; NPQR}_{M}-8\delta_{M}^{N}\Gamma^{PQR}\right]F_{NPQR}\epsilon
\label{KillingSpinEqM2}
\end{equation}
and $F$ is the four-form field strength of 11d supergravity.

It has proven useful to repackage $\epsilon (x)$ in terms of the following one, two and five forms:

\begin{eqnarray}
K_{M} &=& \bar{\epsilon}\Gamma_{M}\epsilon \nonumber \\
\omega_{MN} &=& \bar{\epsilon}\Gamma_{MN}\epsilon \nonumber \\
\Sigma_{MNPQR} &=& \bar{\epsilon}\Gamma_{MNPQR}\epsilon .
\label{SpinorAns}
\end{eqnarray}
Then $\epsilon (x)$ can be reconstructed (up to a sign) from knowledge of $K,\omega$ and $\Sigma$. One can check that the zero, three and four-forms built this way and their duals vanish identically. 

Following the analysis of~\cite{Hackett-Jones:2003vz}, we find that one can rewrite the super-Poincar\'{e} algebra of flat 11d supergravity coupled to a supermembrane probe

\begin{equation}
\left\{ Q_{\alpha},Q_{\beta} \right\} = (C\Gamma^{M})_{\alpha\beta}P_{M} \pm \frac{1}{2}(C\Gamma_{MN})\alpha\beta Z^{MN},
\end{equation}
where the central charge is defined to be

\begin{equation}
Z^{MN} = \int dX^{M}\wedge dX^{N}
\end{equation}
(where the integration is taken over the spatial worldvolume of the membrane) and $Q_{\alpha}$ are the 32 component Majorana spinors charges, in the equivalent way

\begin{equation}
2(\epsilon Q)^2 = K^{M}P_{M} \pm \omega_{MN}Z^{MN}.
\end{equation}

Following~\cite{Hackett-Jones:2003vz}, for a general curved background (without imposing any restriction on $K^{M}$), we have:

\begin{equation}
2(\epsilon Q)^2 = \int K^{M}P_{M} \pm \int (\omega + \imath_{K}A).
\label{M2susygenalgebra}
\end{equation}

In this expression, $A$ is a three-from potential for the four-form field strength $F$. The supersymmetry algebra~(\ref{M2susygenalgebra}) leads to a BPS type bound on the energy momentum of the M2-brane, since $(\epsilon Q)^2 \geq 0$. We find

\begin{equation}
\int K^{M}P_{M} \geq \mp \int (\omega + \imath_{K}A)
\label{BPSboundM2}
\end{equation}
where the term on the RHS is topological in nature.This is indeed the topological bound we shall calculate for our M2-brane in our particular background. 

\subsubsection{Calculation of the BPS bound}

For clarity we restate the supergravity background metric and four-form field strength we shall use for the calculation of the topological objects constructed in the last section. These are

\begin{equation}
ds^2 = H^{-1/3} {dx^2}_{(1,3)} + 2 H^{-1/3} g_{M \bar{N}} dz^{M}dz^{\bar{N}} + H^{2/3} {dx^2}_{(3)}
\end{equation}
where 

\begin{equation}
g_{M \bar{N}} \equiv 2 \left( {\partial}_{M} F^2 \right) \left( \overline{ {\partial}_{N} F^2 } \right) g + 1/2 \left( {\partial}_{M} G \right) \left( \overline{ {\partial}_{N} G } \right)
\end{equation}
with $g= \frac {\pi N}{8 {\tilde{r}}^3 }$ and $\tilde{r} \equiv \sqrt{t^4 + \left\vert F \right\vert ^4}$. \\
The four-form field strength is given by

\begin{eqnarray}
F_{M\bar{N}\alpha\beta} & = & i \epsilon_{\alpha\beta\gamma}\partial_{\gamma}g_{M\bar{N}} \nonumber \\
F_{M89(10)} & = & - 4 i \partial_{M} g \nonumber \\
F_{\bar{N}89(10)} & = & 4 i \partial_{\bar{N}} g .\nonumber 
\end{eqnarray}
The hermitian metric $g_{M\bar{N}}$ can be decomposed into the tangent space zweibeins

\begin{eqnarray}
e^{1}_{M} &=& 2\sqrt{g}\left(\partial_{M}F^2\right) \nonumber \\
\overline{e^{2}_{N}} &=& \left(\overline{\partial_{N}G}\right)
\end{eqnarray}
and so 

\begin{equation}
g_{M\bar{N}} = e^{a}_{M}\left(\overline{e^{b}_{N}}\right)\delta_{a\bar{b}}.
\end{equation}

Now, we saw in the previous subsection that we need to construct the quantity $\omega = \bar{\epsilon}\hat{\Gamma}\epsilon$. We shall need the projection conditions for our particular supersymmetric configuration, which are

\begin{equation}
\hat{\Gamma}_{0123a\bar{b}}\epsilon = i \delta_{a\bar{b}}\epsilon
\label{ProjM5B}
\end{equation}
for the M5-brane, and

\begin{equation}
\left(e^{i\phi}\hat{\Gamma}_{0ab} + e^{-i\phi}\hat{\Gamma}_{0\bar{a}\bar{b}}\right)\epsilon = \epsilon
\label{ProjM2}
\end{equation}
for the M2-brane, where in both cases $\hat{\Gamma}$ denotes the tangent space gamma matrices. We have included an arbitrary phase for the M2-brane projection conditions, which generalises the $\phi =0$ case of Section~\ref{ComplexStructureM2}. We note that the linear combination of holomorphic and anti-holomorphic projection conditions do indeed insure it is an hermitian projector with $\mathcal{P}^2 = 1$. 

From the Killing spinor equation~(\ref{KillingSpinEqM2}), we can also deduce that in fact $$\epsilon (x)=H^{-1/12}\epsilon_{0}$$ (where $\epsilon_{0}$ is a constant spinor). To find all the contributions to the two-form $\omega$ we use the ans\"{a}tze~(\ref{SpinorAns}) and the aforementioned projection conditions~(\ref{ProjM5B},\ref{ProjM2}). So for example, $\omega$ has no contributions of the form $\omega_{0a}$ since

\begin{eqnarray}
\omega_{0a} &=& \bar{\epsilon} {\hat{\Gamma}}_{0a}\epsilon \nonumber \\
& = & -i \bar{\epsilon} {\hat{\Gamma}}_{123a} \epsilon \nonumber \\
& = & 0 \nonumber
\end{eqnarray}
where we have used the fact that $\bar{\epsilon}{\hat{\Gamma}}_{123a}\epsilon$ vanishes identically and also the M5-brane projection conditions. Further, we note that it is possible to view the matrices ${\hat{\Gamma}}_{a}$ and ${\hat{\Gamma}}_{\bar{b}}$ as creation and annihilation operators since we have

\begin{eqnarray}
{\hat{\Gamma}}_{a\bar{b}}{\hat{\Gamma}}_{a} \epsilon & =& \delta_{a\bar{b}} {\hat{\Gamma}}_{a} \epsilon \nonumber \\
{\hat{\Gamma}}_{a\bar{b}}{\hat{\Gamma}}_{\bar{b}} \epsilon & = & - \delta_{a\bar{b}} {\hat{\Gamma}}_{\bar{b}} \epsilon .
\end{eqnarray}

Using these relations, we find that the only non-vanishing components of the two-form $\omega$ give

\begin{equation}
\omega = - H^{-1/6} \left(e^{-i\phi} \epsilon_{\mu\nu} dz^{\mu}\wedge dz^{\nu} + e^{i\phi}\epsilon_{\bar{\mu}\bar{\nu}} d\bar{z^{\mu}} \wedge d\bar{z^{\nu}} \right) .
\end{equation}
Our normalization was chosen such that $\epsilon ^{\mbox{\dag}} \epsilon = H^{-1/6}$. We should also note that the tensors $\epsilon_{\mu\nu}$ include a factor of $H^{-1/3}$ coming in from the warp factor of the metric~(\ref{M2metricN2}).
Now rewriting the above in terms of $w,y$ and $F^2,G$ we find

\begin{eqnarray}
\omega & = & - H^{-1/2} \left( e^{-i\phi} e^{1}_{[1}e^{2}_{2]} dz^{1} \wedge dz^{2} + e^{i\phi} \bar{e}^{1}_{[1}\bar{e}^{2}_{2]} d\bar{z}^{1} \wedge d\bar{z}^{2} \right) \nonumber \\
& = &  1/2 \left( e^{-i\phi} dw \wedge dy + e^{i\phi} d\bar{w} \wedge d\bar{y} \right) = 1/2 \left( e^{-i\phi} dF^2 \wedge dG + e^{i\phi} d\bar{F^2} \wedge d\bar{G} \right) \nonumber \\
& &
\end{eqnarray}
using the fact $H=4g$ in our conventions and also the condition $\left({\partial}_{y}F^{2}\right) \left( {\partial}_{w} G \right) - \left({\partial}_{w}F^{2}\right) \left( {\partial}_{y} G \right) = 1$. 

Finally, we also note that the inner product $\imath_{K}A$ does not give any contributions for our choice of background. This is because the Killing vector $K$ only has non-trivial components in the $(0,1,2,3)$ space, whereas the three-form potential $A$ only has components in the $(4,5,6,7,8,9,10)$ space, so the inner product vanishes.

Therefore, the BPS lower bound on the mass of the M2-brane in our particular background is given by

\begin{equation}
\int K^{M}P_{M} \geq \mp \int \omega 
\end{equation}
in accordance with Equation~(\ref{BPSboundM2}). This also reproduces our earlier result of Equation~(\ref{MassofM2probe}) if we note that under an appropriate worldvolume co-ordinate definition, setting our phase $\phi =0$ and using Equation~(\ref{HolRestraintsFG}) we would have
$$
1/2 \left( dF^2 \wedge dG + d\bar{F^2} \wedge d\bar{G}\right) = dF^2 \wedge dG.
$$

We note that there are no supergravity corrections to the holomorphic two-form which gives the mass of the BPS monopoles in the dual gauge theory. All the Seiberg-Witten analysis then follows through unchanged.

\section{The $\mathcal{N} = 2$ Super Yang-Mills theory}
\label{N2SYM}

We will now show how the supergravity solution~(\ref{FSspacetime}), along with the known form of the M5-brane worldvolume action~(\ref{M5Action}), can be used to extract information about the corresponding gauge theory. We will study the dynamics of an M5-brane probe which wraps around the M-theory direction, and thus reduces to a D4-brane upon dimensional reduction. We will also calculate the Yang-Mills coupling and the theta angle for the $\mathcal{N} = 2$ gauge theory living on the D4-brane worldvolume. For a similar analysis in the Type IIB picture see~\cite{DiVecchia:2002ks}.

\subsection{Reduction process for the M5-brane worldvolume action}

The first step is to dimensionally reduce the M5-brane worldvolume action along the M-theory direction to arrive at the D4-brane action. This is actually a two-step process, as a direct dimensional reduction yields the so-called dual D4-brane action. So after performing the reduction, we then have to dualise the resulting action to arrive at the usual string frame DBI action for the D4-brane. We will use and follow the analysis of~\cite{Aganagic:1997zk} for these steps, and refer the reader there for further details. 

It is important to note that we will be using a modified Kaluza-Klein reduction ans\"{a}tze. Explicitly, the eleven dimensional metric can be expressed in component form as 

\begin{equation}
G_{\hat{\mu}\hat{\nu}} = \left( \begin{array}{cc} e^{-2\phi/3}\left(g_{\mu \nu} + C_{\mu} C_{\nu} \right) & \upsilon e^{\phi/3} C_{\mu} \\
\upsilon e^{\phi/3} C_{\nu} & \upsilon^{2} e^{4\phi/3} \end{array} \right)
\label{KKmatrix}
\end{equation}
where $\upsilon$ is the winding number, giving the number of times the M5-brane wraps the compactified dimension and $C_{\mu}$ is the R-R one-form. For the M5-brane worldvolume reduction we shall set $\upsilon = 1$. We can rewrite the M5-brane action in the form

\begin{eqnarray}
S_{M5}  =  -\tau_{5} \int d^6\sigma \left[ \sqrt{-G_{6}}\sqrt{1+\hat{z}_{1}  
+ \hat{z}_{1}^{2}/2 - \hat{z}_{2}}  \right. \nonumber \\
 \left. + \frac{1}{24(\partial a)^2}{\epsilon}^{\hat{l}\hat{m}\hat{n}\hat{p}\hat{q}\hat{r}} \hat{\mathcal{H}}_{\hat{p}\hat{q}\hat{r}} \hat{\mathcal{H}}_{\hat{m}\hat{n}\hat{s}} g^{\hat{s}\hat{t}} {\partial}_{\hat{l}}a {\partial}_{\hat{t}}a + {\mathcal{L}}_{WZ} \right]
\label{M5reduxaction}
\end{eqnarray}
where we have denoted the 6d (worldvolume) coordinates by $\sigma^{\hat{\mu}}=(\sigma^{\mu},\sigma^{5})$ with $\mu=0,1,2,3,4$ and $G_{6}$ is the 6d determinant. The $z$ variables are defined to be

\begin{eqnarray}
\hat{z}_{1} &=& \frac{\mbox{tr}(\hat{\tilde{\mathcal{H}}}^2)}{2} = \frac{G_{\hat{\mu}\hat{\nu}}\hat{\tilde{\mathcal{H}}}^{\hat{\nu}\hat{\rho}}G_{\hat{\rho}\hat{\lambda}}\hat{\tilde{\mathcal{H}}}^{\hat{\lambda}\hat{\mu}}}{2} \\ 
\hat{z}_{2} &=& \frac{\mbox{tr}(\hat{\tilde{\mathcal{H}}}^4)}{4} =  \frac{\hat{G}\hat{\tilde{\mathcal{H}}}\hat{G}\hat{\tilde{\mathcal{H}}}\hat{G}\hat{\tilde{\mathcal{H}}}\hat{G}\hat{\tilde{\mathcal{H}}}}{4}.
\end{eqnarray}
If we now fix the gauge so that the compactified direction is taken to be the $a=\sigma^{5}$ direction (and hence $\partial_{\hat{\mu}} a = \delta^{5}_{\hat{\mu}}$), then we find that the quantity 

\begin{equation}
 (\partial a)^2  = G^{\hat{\mu}\hat{\nu}}\partial_{\hat{\mu}}a\partial_{\hat{\nu}}a
\end{equation}
reduces to $G^{55}$, and both $G^{55}$ and $G^{\rho5}$ are components of the 6d inverse metric $G^{\mu\nu}$. As direct calculation shows, these are given by $G^{55}=(1+C^2)$ and $G^{\rho5}=-e^{\phi}C^{\rho}$.

Upon dimensional reduction, the second term in the M5-brane action~(\ref{M5reduxaction}) above splits into two, in particular

\begin{equation}
{\mathcal{L}}_{\mathcal{H}\tilde{\mathcal{H}}} \longrightarrow \tau_{5}\left(-\frac{\epsilon_{\mu\nu\lambda\sigma\tau}}{8(1+C^2)}e^{\phi}C^{\mu}\tilde{\mathcal{H}}^{\nu\lambda}\tilde{\mathcal{H}}^{\sigma\tau} + \frac{1}{24}\epsilon^{\mu\nu\rho\lambda\sigma(5)}\mathcal{H}_{\rho\lambda\sigma} \mathcal{H}_{\mu\nu(5)} \right)
\end{equation}
where now the second term above contributes to the original Wess-Zumino term to form a new term $WZ'$ (see~\cite{Aganagic:1997zq} for more details). We have used the explicit expressions for $G^{55}$ and $G^{\mu5}$ in the first term. 

Fixing the gauge and dimensionally reducing the $DBI$ term yields 

\begin{equation}
\mathcal{L}_{DBI} \longrightarrow \tau_{5} e^{-\phi}\sqrt{-G_{5}}\sqrt{1+e^{2\phi}z_{1}  
+ e^{4\phi}(z_{1}^{2}/2 - z_{2})}
\end{equation}
where 

$$
\begin{array}{rcl}
z_{1} & = &  \frac{1}{2} \mbox{tr} \left(\tilde{\mathcal{H}}^2\right) \nonumber \\
z_{2} & = & \frac{1}{4} \mbox{tr} \left(\tilde{\mathcal{H}}^4\right)
\end{array}
$$
and the dimensional reduction of the field $\hat{\tilde{\mathcal{H}}}$ is given by the expression:

\begin{equation}
\hat{\tilde{\mathcal{H}}} \rightarrow e^{\frac{10}{3}\phi} \tilde{\mathcal{H}}.
\end{equation}
We have used the fact that  $\hat{z}_{1} \rightarrow e^{2\phi}z_{1}$ and $\hat{z}_{2} \rightarrow e^{4\phi}z_{2}$ 

We will now rescale the $\Phi$ and $\mathcal{H}$ fields to absorb the factor of the M5-brane tension $\tau_{5}$ in front of the $DBI$ and $\mathcal{H}\tilde{\mathcal{H}}$ term. This will then put our action in the same form as that of~\cite{Aganagic:1997zk}, and their dualization procedure follows trivially. The rescalings are of the form

\begin{eqnarray}
e^{\phi'}&=&e^{\phi-\ln\tau_{5}} \\
\tilde{\mathcal{H}}' &=& \tau_{5}\tilde{\mathcal{H}},
\end{eqnarray}
so we see that the combinations $e^{2\phi}z_{1}$ and $e^{4\phi}z_{2}$ appearing inside the square root above are actually invariant under this rescaling.

So grouping together the various terms we can rewrite our compactified M5-brane worldvolume action, which is actually the dual D4-brane action, as

\begin{eqnarray}
S^{*}_{D4} &=& -  \int d^5\sigma \left( e^{-\phi'}\sqrt{-G_{5}}\sqrt{1+e^{2\phi'}z_{1}'  
+ e^{4\phi'}(z_{1}'^{2}/2 - z_{2}')} \right. \\ 
& &  \left. + \frac{\epsilon_{\mu\nu\lambda\sigma\tau}}{8(1+C^2)}e^{\phi'}C^{\mu}\tilde{\mathcal{H}}'^{\nu\lambda}\tilde{\mathcal{H}}'^{\sigma\tau}  \right) + \int \mathcal{L}_{WZ'}
\end{eqnarray}
with the modified Wess-Zumino term given by

\begin{equation}
WZ' = e^{-\phi'}C^{(5)} + \frac{1}{2}{\mathcal{\tilde{H}}}'\wedge C^{(3)} + \frac{1}{4}{\mathcal{\tilde{H}}}'^{\mu\nu}\partial_{(5)}B_{\mu\nu} .
\end{equation}
It now follows from the analysis of~\cite{Aganagic:1997zk} that the D4-brane action with a constant dilaton background field is given by

\begin{eqnarray}
S_{D4} &=& -  \int d^5\sigma \; e^{-\phi'}\sqrt{-\mbox{det}(G_{\mu\nu} + \mathcal{F}_{\mu\nu})} \\
& & -  \int e^{-\phi'} \left( C_{(5)} + C_{(3)}\wedge\mathcal{F} + \frac{1}{2}C_{(1)}\wedge\mathcal{F}\wedge\mathcal{F}\right) .
\label{D4Action}
\end{eqnarray}
The two are related by  

\begin{equation}
-\frac{\delta S_{D4}}{\delta F_{\mu\nu}} = \tilde{H}^{\mu\nu}
\label{DualHF}
\end{equation}

where we note that we have the 6d hodge duals $\tilde{H}=\ast H$ and the definitions $\hat{\mathcal{H}} \equiv H - e^{-\phi}C^{(3)}$ and $\mathcal{F} = F - b_{(2)}$. The method of~\cite{Aganagic:1997zk} relies on constructing Lorentz invariant quantities with a particularly simple choice for the form of $F_{\mu\nu}$, which is then used to solve Eq.~(\ref{DualHF}). Since the quantities are Lorentz invariant, it is straightforward to pass from this special frame to a more general frame.

\subsection{Dimensional reduction of the background supergravity solution}

Before we proceed any further, we also need to dimensionally reduce the background supergravity solution~(\ref{FSspacetimeFG}) down to the ten-dimensional Type IIA string frame metric. We recall that we are using a modified Kaluza-Klein reduction ans\"{a}tze which, expressed as a line element, has the form

\begin{eqnarray}
ds^{2}_{(1,10)} & = & e^{-2\phi/3} ds^{2}_{(1,9)} + e^{4\phi/3} \left(dx^{7} + e^{-\phi} C_{\mu}dx^{\mu} \right)^{2} \\ 
\hat{F}_{(4)} & = & \mathcal{F}_{(4)} + \mathcal{T}_{(3)}\wedge dx^{7}
\end{eqnarray}
where $\hat{F}_{(4)}=dC_{(3)}$ is the field strength for the background three-form potential $C_{(3)}$, with $\mathcal{F}_{(4)}$ and $\mathcal{T}_{(3)}=dI_{(2)}$ being the RR four-form and the NSNS three-form field strengths of the ten dimensional Type IIA theory. We recall that the coordinate $x^{7}$ is the circle (of radius $R$) we are compactifying on, with periodicity $2\pi R$. 

For clarity and ease of reading, we write down the eleven dimensional supergravity solution we stated earlier.

\begin{equation}
ds^2 = H^{-1/3} {dx^2}_{(1,3)} + 2 H^{-1/3} g_{M \bar{N}} dz^{M}dz^{\bar{N}} + H^{2/3} {dx^2}_{(3)}
\end{equation}
with $M,N=F^{2},G$ and  where we recall that $F^{2}(w,y)$ and $G(w,y)$ are holomorphic functions of $w,y$ with

$$
\begin{array}[c]{ccccc}
w & = & \frac{v}{{l_{s}}^{2}} & = & \frac{vR}{{l_{P}}^{3}} \\
t^{2} & = & \frac{r}{g_{S} {l_{s}}^{3}} & = & \frac{r}{{l_{P}}^{3}} \nonumber \\
y & = & \frac{s}{R} & &   \\
\end{array}
$$
\\
and 

\begin{eqnarray}
v & = & x^{4} + i x^{5} \nonumber \\
s & = & x^{6} + i x^{7}.
\end{eqnarray}

\begin{figure}[htb]
\centering
\epsfig{file=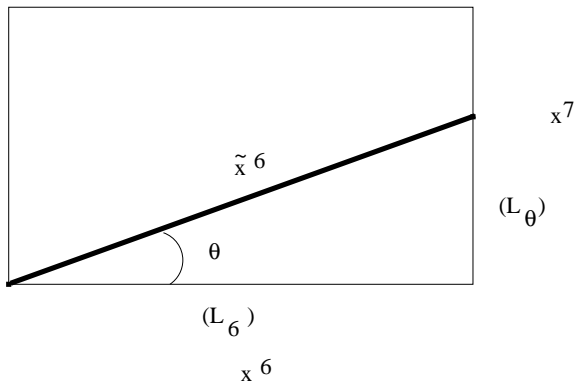}
\caption{The tilted D4-brane with endpoints at different values of $x^7$.}
\label{fig:x6x7theta}
\end{figure}

We can rewrite the complex hermitian metric $g_{M \bar{N}}$ in terms of real coordinates, and then use the hermiticity condition to simplify it further. Our aim is to calculate field theory quantities on the resulting D4-brane worldvolume action, in which case the endpoints of the D4-brane are allowed to have different fixed values in the $x^{7}$ direction. With this in mind, we modify our original coordinates

\begin{eqnarray}
\tilde{x}^{6} & = & x^{6}\cos \theta  + x^{7}\sin \theta  \nonumber \\
\tilde{x}^{7} & = & x^{7}
\end{eqnarray}
so that we are effectively tilting in the $x^{7}$ direction (see Fig.~\ref{fig:x6x7theta}). We consider only constant $\theta$ values.

The resulting 11-d metric is then given by

\begin{equation}
ds^{2}_{11} = H^{-1/3}dx^{2}_{(1,3)} + 2H^{-1/3}M_{\mu\nu}dx^{\mu}dx^{\nu} + 2H^{-1/3}ds^{2}_{KK} + H^{2/3}dx^{2}_{\perp}
\label{11dKKmetric}
\end{equation}
with the Kaluza-Klein part of the metric being

\begin{equation}
ds^{2}_{KK} = \frac{g_{7\bar{7}}}{\cos^{2}\theta}\left( d\tilde{x}^{7} + D_{\mu}dx^{\mu} \right)^{2} .
\end{equation}
If we denote the $\mu\nu=(a,b,\tilde{6})$, then the $M_{\mu\nu}$ part corresponds to 

\begin{eqnarray}
M_{\mu\nu}dx^{\mu}dx^{\nu} & = & \left( g_{a\bar{b}} - \frac{\cos^2\theta k_{a}k_{b}}{2g_{7\bar{7}}} \right) dx^{a}dx^{b} - \frac{\cos^2\theta k_{a}^2}{ 4 g_{7\bar{7}}} dx^{a}d\tilde{x}^{6} \nonumber \\
& & + \sin\theta k_{b}d\tilde{x}^6dx^{b} + \left( g_{\tilde{6}\tilde{\bar{6}}}  - g_{6\bar{6}}\tan^2\theta\right)d\tilde{x}^6d\tilde{x}^6
\label{MpartMetric}
\end{eqnarray}
where $k_{a}=\left[ (g_{a\bar{7}} + g_{7\bar{a}}) - \tan\theta(g_{a\bar{6}} + g_{6\bar{a}})\right]$, with $a,b=4,5$. The notation for $g_{7\bar{7}}$ refers to 

\begin{equation}
g_{7 \bar{7}} = 2g \left( {\partial}_{7} F^2 \right) \left(  {\partial}_{7} \bar{F^2}  \right)  + 1/2 \left( {\partial}_{7} G \right) \left(  {\partial}_{7} \bar{G}  \right)
\end{equation}
and $D_{\mu}$ is in general given by

\begin{equation}
D_{\mu}dx^{\mu} = \frac{\cos^{2}\theta}{2g_{7\bar{7}}}\left[ (g_{a\bar{7}} + g_{7\bar{a}}) - \tan\theta(g_{a\bar{6}} + g_{6\bar{a}})\right]dx^{a} -\sin\theta d\tilde{x}^{6}.
\end{equation}

\subsection{The Yang-Mills coupling and theta angle}

Having arrived at the expression for the D4-brane action~(\ref{D4Action}), we now show how the supergravity solution~(\ref{FSspacetimeFG}) can be used to extract information about the corresponding gauge theory. 

We wish to study the dynamics of the $SU(N)$ gauge fields that propagate on the wrapped M5-branes. We will look at Hanany-Witten type configurations where the D4-brane is finite in extent in the $\tilde{x}^6$ direction, which effectively reduces the worldvolume degrees of freedom to four. This four-dimensional part is flat, and is the Minkowski space-time $\mathbf{R}^{(1,3)}$ where a supersymmetric gauge theory with eight supercharges is defined. One finds that the low-energy four-dimensional gauge theory is a pure $\mathcal{N}=2$ SYM theory with gauge group $SU(N)$. 

The action for our M5-brane probe reduced to a D4-brane was calculated in the last section to be

\begin{eqnarray}
S_{D4} &=& -  \int d^5\sigma \; e^{-\phi'}\sqrt{-\mbox{det}(G_{mn} + \mathcal{F}_{mn})} \\
& & -  \int e^{-\phi'} \left( C_{(5)} + C_{(3)}\wedge\mathcal{F} + \frac{1}{2}C_{(1)}\wedge\mathcal{F}\wedge\mathcal{F}\right)
\end{eqnarray}
where all the bulk fields are understood to be the pull-backs onto the brane world-volume which is parametrised by $\sigma=\left( x^{0},x^{1},x^{2},x^{3},\tilde{x}^{6} \right)$. By expanding the square root part of the above action and examining the component which is quadratic in the field strength $\mathcal{F}$ we can deduce the Yang-Mills coupling for the world-volume theory. 
We achieve this by promoting $\mathcal{F}$ to an $SU(N)$ field and by giving it an adjoint index $A$. If the generators are normalized in such a way that $\mbox{tr}\left(T^{A}T^{B}\right)=(1/2)\delta^{AB}$ for the fundamental representation, then the above procedure leads to 

\begin{equation}
S_{YM} = -\frac{1}{g^{2}_{YM}}\int d^{4}\sigma \frac{1}{4} \mathcal{F}^{A}_{\alpha\beta}\mathcal{F}^{\alpha\beta}_{A} + \frac{\left(\Theta_{YM} + 2\pi n \right)}{32\pi^2} \int d^4\sigma \mathcal{F}^{A}_{\alpha\beta}\tilde{\mathcal{F}}^{\alpha\beta}_{A},
\end{equation}
where 

\begin{equation}
\frac{1}{g^{2}_{YM}} = \frac{\tau_{5}}{2} \left(2\pi\alpha'\right)^2 \int d\tilde{x}^{6} \; d\tilde{x}^{7} \; e^{-\phi} \sqrt{-G_{5}}\left(g^{mn}\right)^2
\end{equation}
for $(m,n)=0,1,2,3$ (i.e. $\left(g^{mn}\right)^2$ refers to two factors of the inverse metric in the $0,1,2,3$ flat space) and $\sqrt{-G_{5}}$ denotes the square root determinant of the induced worldvolume metric. 

We can also deduce the value of the Yang-Mills theta angle to be

\begin{equation}
\Theta_{YM} = \left[ -\tau_{5} (2\pi)^2 \left(2\pi\alpha'\right)^2\int d\tilde{x}^{6} \; d\tilde{x}^{7} \; e^{-\phi} C^{(1)}_{6} \right] \mod 2\pi
\end{equation}
where $C^{(1)}_{6}$ denotes the six component of the one form $C^{(1)}$. 

Now, to calculate the pull-back of the various forms we need to establish the ten-dimensional background metric in the string frame. We can do this by comparing our 11-d metric~(\ref{11dKKmetric}) with our Kaluza-Klein reduction ans\"{a}tze metric

\begin{equation}
ds^{2}_{(1,10)}  =  e^{-2\phi/3} ds^{2}_{(1,9)} + e^{4\phi/3} \left( d\tilde{x}^{7} + e^{-\phi} C_{\mu}dx^{\mu} \right)^{2} 
\label{10dMetric}
\end{equation}
for $\mu=4,5,6$. Due to the hermitian nature of the metric, the $D_{6}$ component of the 11d metric simplifies to

\begin{equation}
D_{6} = -\sin\theta
\end{equation}
where $D_{6}=g^{1/4}\left[\frac{8g_{7\bar{7}}}{\cos^6\theta}\right]^{-1/4}C_{6}$.

We can then read off the R-R ($C_{\mu} $) and NS-NS ($\phi , g_{\mu\nu}$) fields from the dimensional reduction of the background metric~(\ref{11dKKmetric}). We find the dilaton is given by

\begin{equation}
\phi = \frac{3}{4} \ln \left( \frac{2g_{7\bar{7}}}{ \cos^2 \theta g^{1/3}} \right) ,
\label{10dDilaton}
\end{equation}
and for the six component of the R-R one form we get

\begin{equation}
C_{6}^{(1)} = -g^{-1/4}\sin\theta\left[\frac{8g_{7\bar{7}}^3}{\cos^6\theta}\right]^{(1/4)} .
\label{10dOneForm}
\end{equation}
The ten-dimensional string frame background metric is given by

\begin{equation}
ds^{2}_{10} = \frac{\sqrt{2g_{7\bar{7}}}}{\cos\theta}\left[H^{-1/2}dx^{2}_{(1,3)} + 2H^{-1/2}M_{\mu\nu}dx^{\mu}dx^{\nu} + H^{1/2}dx^{2}_{\perp}\right] .
\end{equation}
If we now place our probe so that it lies along the $0123[\tilde{6}]$ directions, we can calculate the induced metric and thus the Yang-Mills coupling. If we take into account the possibility of our probe wrapping the $x^7$ direction $N$ times, this corresponds to looking at an SU($N$) gauge theory instead of a U(1) gauge theory. The coupling turns out to be 

\begin{equation}
\frac{1}{g^2_{YM}}= \frac{N}{8\pi^2 g_{s}l_{s}} \int d\tilde{x}^{6}  \;  \cos\theta
\label{YMcoupling}
\end{equation}
where to evaluate the determinant for the induced worldvolume metric we have used the results $g_{\tilde{6}\bar{\tilde{6}}}=g_{6\bar{6}}/\cos^2\theta$, $g_{6\bar{6}}=g_{7\bar{7}}$ and $M_{\tilde{6}\tilde{6}}=g_{7\bar{7}}$, which follow from the hermitian condition of the metric components.  The theta angle turns out to be quite simple as well, explicitly

\begin{equation}
\Theta_{YM} = \left[ \frac{N}{g_{s}l_{s}} \int d\tilde{x}^{6}  \; (  \sin\theta) \right] \mod 2\pi N .
\label{YMthetaangle}
\end{equation}
So the end result is that the usual Seiberg-Witten analysis goes through unchanged. The bending of the NS5-branes given by the Seiberg-Witten curve is encoded in the $\tilde{x}^6$ integral. 

For the canonical example of the Hanany-Witten Type IIA model with two NS5-branes separated by a distance $L_{6}$ and the D4-brane endpoints on the NS5-branes separated in the $x^7$ direction by a distance $L_{\theta}$, the gauge coupling and theta angle reduce to the classical values

\begin{eqnarray}
\frac{1}{g^{2}_{YM}} & = & \frac{L_{6}}{8\pi^2g_{s}l_{s}} \nonumber \\
\Theta_{YM} & = & \frac{L_{\theta}}{g_{s}l_{s}} \nonumber
\end{eqnarray}
and the $\mathcal{N}=2$ complex gauge coupling can be written in the usual form

\begin{eqnarray}
\tau & = & \frac{\Theta_{YM}}{2\pi} + i \frac{4\pi}{g^{2}_{YM}} .
\end{eqnarray}

\subsection{Instantons}

We can also show that instantons are correctly represented as Euclidean D0-branes living on the colour D4-branes~\cite{Barbon:1997ak,Brodie:1998bv} in the $\tilde{x}^6$ direction. The world-volume action of a Euclidean D0-brane in our special frame is given by

\begin{equation}
S_{D0} = \tau_{0} \int d\tilde{x}^6 \; e^{-\phi}\sqrt{G_{\tilde{6}\tilde{6}}} - i \tau_{0} \int e^{-\phi}C^{(1)}_{6}
\end{equation}
where 

\begin{equation}
\tau_{0} = \frac{1}{g_{s}l_{s}},
\end{equation}
and the bulk fields are understood to be the pullbacks onto the brane worldvolume. The appearance of $i$ is due to the Wick rotation we perform to arrive at the Euclidean action.

Using the ten-dimensional metric we computed earlier~(\ref{10dMetric}), as well as the dilaton~(\ref{10dDilaton}), the R-R one-form~(\ref{10dOneForm}) and the explicit form of the metric~(\ref{MpartMetric}) it is easy to see that

\begin{equation}
S_{D0} = \frac{8\pi^2}{g^2_{YM}} - i\Theta_{YM}
\end{equation}
which is the correct form of the instanton action. We have used the previous expressions for the Yang-Mills coupling~(\ref{YMcoupling}) and theta angle~(\ref{YMthetaangle}) to arrive at this result. So we conclude that the gauge theory instantons of the $\mathcal{N}=2$ SYM theory are indeed represented by Euclidean D0-branes extended in the $\tilde{x}^6$ direction, as one should expect from general considerations.

\section{The $\mathcal{N} = 1$ Super Yang-Mills theory}
\label{N1SYM}

The previous construction, which describes the eleven-dimensional supergravity dual of $\mathcal{N}=2$ field theories as the near-horizon limit of an M5-brane wrapped on a Riemann surface $\Sigma \subset \mathbf{C}^2$, has been generalised to $\mathcal{N}=1$. In particular, the eleven dimensional supergravity dual of certain $\mathcal{N}=1$ field theories (so-called MQCD theories~\cite{Hori:1997ab,Witten:1997ep}) is given by the near-horizon limit of an M5-brane wrapped on a Riemann surface $\Sigma \subset \mathbf{C}^3$. MQCD is then the quantum field theory living on the $(0,1,2,3)$ part of a five-brane with world-volume $(0,1,2,3) \times \Sigma$.

\subsection{The M5-brane configuration}

The idea is very similar to the $\mathcal{N}=2$ case, where we begin with a system of NS5-branes and D4-branes in type IIA string theory. As an illustration, we can look at the simplest case of pure Yang-Mills with no matter. This is realised by two NS5-branes, denoted by NS$5_{1}$ and NS$5_{2}$. The NS$5_{1}$ brane has world-volume directions $012345$, while the NS$5_{2}$ brane has world-volume directions $012389$. They are separated in the $6$ direction with the NS$5_{1}$ brane defined to be on the left. We can then consider the inclusion of $n$ D4-branes of finite extent in the $6$ direction which are suspended between the NS5-branes. This configuration will then describe an $\mathcal{N}=1$ four-dimensional SU($n$) field theory on the world-volume of the $n$ finite D4-branes.

These configurations can be lifted to M-theory where they become an M5-brane wrapped on a non-compact Riemann surface $\Sigma$ embedded in $\mathbf{C}^3$, generalising the $\mathcal{N}=2$ case of $\Sigma \subset \mathbf{C}^2$.

This is also equivalent to starting with the $\mathcal{N}=2$ configuration of Section~\ref{N2SYM} and rotating one of the NS5-branes from the 45 plane onto the 89 plane. This corresponds to turning on a mass for the adjoint scalar in the $\mathcal{N}=2$ vector multiplet, breaking the supersymmetry to $\mathcal{N}=1$. More general setups describing field theories with different gauge groups and matter have been constructed (see for instance~\cite{Giveon:1998sr} and related papers). For a similar anlaysis from the Type IIB viewpoint see also~\cite{DiVecchia:2002ks,Nunez:2003cf}.

\subsection{The supergravity solution}

The supersymmetry preserving solutions of eleven-dimensional supergravity relevant for describing the M5-brane setup were described in \cite{Brinne:2000nf}. The method is very similar to the $\mathcal{N}=2$ case, so we shall go straight to the results. The solution was found to be:

\begin{eqnarray}
ds^2 &=& H^{-1/3} {dx^2}_{(1,3)} + 2 H^{1/6} g_{M \bar{N}} dz^{M}dz^{\bar{N}} + H^{2/3} dy^2  
\label{N1Metric} \\
\det g &=& H \nonumber \\
F &=& \partial_{y}(\omega\wedge\omega)-i\partial(H^{1/2}\omega)\wedge dy + i\bar{\partial}(H^{1/2}\omega)\wedge dy \\
\bar{\partial}(\omega\wedge\omega) &=& 0 \; = \; \partial(\omega\wedge\omega) .
\label{N1omwedgeomconstraint}
\end{eqnarray}
In the above equations, the $z^{M}$ are holomorphic coordinates:

$$
\begin{array}{ccccc}
z^1 &=& v &=& x^4 + ix^5  \\
z^2 &=& w &=& x^6 + ix^7 \\
z^3 &=& s &=& x^8 + ix^9 .
\end{array}
$$
The metric~(\ref{N1Metric}) is of the form $\mathbf{R}^{(1,3)} \times \Sigma \times \mathbf{R}^{(1)}$, where $\Sigma \subset \mathbf{C}^3$ and $y$ denotes the remaining totally transverse direction. Also, $\partial$ denotes the $(1,0)$ exterior derivative $\partial = dz^M\partial_{M}$ in $\mathbf{C}^3$. The metric tensor $g_{M\bar{N}}$ is hermitian, a property we shall use in the following calculations. It has an associated hermitian 2-form

\begin{equation}
\omega = i g_{M\bar{N}} dz^M\wedge dz^{\bar{N}}
\end{equation}
which is useful in expressing the field strength $F$ in a more elegant form. One can check that the $\mathcal{N}=2$ solution satisfies the above constraints.

\subsubsection{An alternative method}
\label{AltSugraSol}

An alternative and fairly straightforward method of finding the supergravity solution is applying ideas from the recent work on the classification of supersymmetric solutions of 11d supergravity~\cite{Gauntlett:2002fz} (see also~\cite{Martelli:2003ki}). We shall demonstrate a derivation of the supergravity solution using these ideas.

Firstly, it is useful to realise that the spinor $\epsilon(x)$ which satisfies the Killing spinor equation~(\ref{KillingSpinEq}) can be reconstructed (up to a sign) from the following one, two and five-forms:

\begin{eqnarray}
K_{M} & = &   \overline{\epsilon}\Gamma_{M}\epsilon \nonumber \\
\Omega_{MN} & = & \overline{\epsilon}\Gamma_{MN}\epsilon   \nonumber \\
\Sigma_{MNPQR} & = & \overline{\epsilon}\Gamma_{MNPQR}\epsilon .
\label{Onetwofiveforms}
\end{eqnarray}
One can check that the zero, three and four-forms built in a similar way vanish identically. 

Furthermore, as recent work on G-structures and related ideas has brought to the fore, if we start with a D=11 geometry with a $Spin(10,1)$ structure and assume that we have a globally defined spinor, then, at a point, the isotropy group of the spinor is known to be either $SU(5)$ or $(Spin(7) \ltimes {\mathbb{R}}^{8}) \times \mathbb{R}$ depending on whether $K$ is time-like or null, respectively. 

Using the fact that our particular background preserves four Killing spinors, we can always consider the case where $K$ is null. The forms $K,\Omega$ and $\Sigma$ then define a $(Spin(7) \ltimes {\mathbb{R}}^{8}) \times \mathbb{R}$ structure corresponding to the stability group of the spinor $\epsilon$. A possible set of tangent space projection conditions for the spinor $\epsilon(x)$ is given in this case by:

\begin{equation}
\hat{\Gamma}_{0123z_{a}\bar{z_{b}}} \epsilon =  i \delta_{a\overline{b}}\epsilon .
\label{ProjM5Spin}
\end{equation}
However, for explicit calculations,  we can choose an arbitrary spinor satisfying this constraint by choosing further appropriate projection conditions. A compatible projection choice along the $01$ directions would be

\begin{equation}
\hat{\Gamma}_{01}\epsilon  =  \pm \epsilon
\end{equation}
where the ambiguity of sign comes from the requirement that the projector squares to 1. It should be emphasised that the equations for $1/8$-SUSY hold for arbitrary $\pm$ sign of this projection, all that is required is that Equation~(\ref{ProjM5Spin}) is satisfied. Similarly, for the $z_{1},z_{2},z_{3}$ space, we may use this freedom to choose the compatible projection condition

\begin{equation}
1/8 \left( e^{i\phi}\hat{\Gamma}_{2z_{1}z_{2}z_{3}} + e^{-i\phi}\hat{\Gamma}_{2\overline{z_{1}z_{2}z_{3}}} \right)\epsilon  =  \epsilon
\end{equation}
and we can check this projector is also hermitian, as is required. Again, the equations for $1/8$-SUSY hold for arbitrary phase $\phi$.

Note that using the identity $\hat{\Gamma}_{0123456789y}\equiv 1$ we can show that our projections imply:

\begin{equation}
\hat{\Gamma}_{y}\epsilon = -\epsilon .
\label{GammayProj}
\end{equation}
These provide a set of independent, commuting projections which determine a unique spinor up to scale. The scale of the spinor is given by fixing 

\begin{equation}
\epsilon^{\dagger}\epsilon  = \Delta.
\end{equation}
We will determine the value of $\Delta$ shortly. 

To calculate the forms and solve the differential equations for the field strength, we first need to determine the form of the metric. We shall start with a metric ans\"{a}tze of the form $\mathbf{R}^{(1,3)} \times \mathbf{C}^3 \times \mathbf{R}^{(1)}$ with the assumption that the complex space is hermitian. This is typical of M5-branes wrapping 2-cycles in $\mathbf{C}^3$. In general we can have

\begin{equation}
ds^{2}=f^2dx_{(1,3)} + 2 e^{a}_{M}\left(\overline{e^{b}_{N}}\right)\delta_{a\overline{b}} dz^{M}dz^{\overline{N}}+ g^2dy^2
\end{equation}
with $f,g$ arbitrary functions of $z^{M},z^{\overline{N}}$ and $y$. As before we let $a,b$ run through $1,2,3$ and normalise the complex part of the metric such that $\delta_{1\overline{1}}=1/2$.

We can now proceed to calculate the non-trivial components of each form. A quick calculation reveals that the $K_{i} (i=2,3), K_{a}, K_{\overline{b}} (a,b=1,2,3)$ and $K_{y}$ components vanish since, for example,
$$
\begin{array}{ccccccccc}
\hat{K}_{y} & = & \overline{\epsilon} \hat{\Gamma}_{y} \epsilon & = & - \overline{\epsilon} \left(\hat{\Gamma}_{y}\right)^{2} \epsilon & = & - \overline{\epsilon} \epsilon & = & 0 
\end{array}
$$
where in the second step we have used the $\hat{\Gamma}_{y}$ projection condition~(\ref{GammayProj}), and in the last step the fact that $\overline{\epsilon} \epsilon$ vanishes identically.

The value of $\Delta$ can be determined taking advantage of the fact that $K$ is a Killing vector. A brief calculation shows that this Killing vector is given by

\begin{equation}
K = - \Delta f \left( dt \mp dx^1 \right).
\end{equation}
Its defining property is that the Lie derivative of our metric ans\"{a}tze with respect to this vector should vanish. This yields a number of constraints which collectively imply

\begin{equation}
\partial_{q} \left( \frac{f}{\Delta} \right) = 0 \Longrightarrow \frac{f}{\Delta} = constant
\end{equation}
with $q$ running over all the spacetime co-ordinates $(0 \ldots y)$. In our normalisation we set this constant equal to one which fixes the value of $\Delta$ to be

\begin{equation}
\Delta = f = \sqrt{-g_{00}}.
\end{equation}

We can proceed in a similar fashion to determine the non-trivial components of the two and five-forms. From our metric ans\"{a}tze we can compute the relevant components. We find

\begin{eqnarray}
K & = & -f^2 \left( dt \mp dx^1 \right) \\
\Omega &=& f^2 g \left( dt \mp dx^1 \right) \wedge dy \\
\Sigma &=& \mp i f^4 e^{a}_{[M}e^{\bar{b}}_{\bar{N}]} \delta_{a\bar{b}} \left(dt \mp dx^1 \right) \wedge dx^2 \wedge dx^3 \wedge dz^{M} \wedge dz^{\bar{N}} \nonumber \\
& & -f^2 e^{a}_{[M}e^{\bar{b}}_{\bar{N}}e^{c}_{P}e^{\bar{d}}_{\bar{Q}]} \delta_{a\bar{b}}\delta_{c\bar{d}} \left(dt \mp dx^1 \right) \wedge dz^{M} \wedge dz^{\bar{N}} \wedge dz^{P} \wedge dz^{\bar{Q}} \nonumber \\
& & + \frac{e^{-i\phi}}{16} f^3 e^{1}_{[1}e^{2}_{2}e^{3}_{3]}  \left(dt \mp dx^1 \right) \wedge dx^2 \wedge dz^1 \wedge dz^2 \wedge dz^3 \nonumber \\
& & + \frac{e^{i\phi}}{16} f^3 e^{\bar{1}}_{[\bar{1}}e^{\bar{2}}_{\bar{2}}e^{\bar{3}}_{\bar{3}]}  \left(dt \mp dx^1 \right) \wedge dx^2 \wedge dz^{\bar{1}} \wedge dz^{\bar{2}} \wedge dz^{\bar{3}} \nonumber \\
& & \pm i \frac{e^{-i\phi}}{16} f^3 e^{1}_{[1}e^{2}_{2}e^{3}_{3]}  \left(dt \mp dx^1 \right) \wedge dx^3 \wedge dz^1 \wedge dz^2 \wedge dz^3 \nonumber \\
& & \mp i \frac{e^{i\phi}}{16} f^3 e^{\bar{1}}_{[\bar{1}}e^{\bar{2}}_{\bar{2}}e^{\bar{3}}_{\bar{3}]}  \left(dt \mp dx^1 \right) \wedge dx^3 \wedge dz^{\bar{1}} \wedge dz^{\bar{2}} \wedge dz^{\bar{3}} .
\end{eqnarray}
Now $\epsilon(x)$ being a Killing spinor also implies that $K, \Omega$ and $\Sigma$ satisfy a set of differential equations. These were given in~\cite{Gauntlett:2002fz}:

\begin{eqnarray}
\label{DiffK}
dK & = &  \frac{2}{3} \iota_{\Omega}F + \frac{1}{3}\iota_{\Sigma}\ast F \\
\label{DiffOmega}
d\Omega  & = &  \iota_{K}F \\
d\Sigma & = & \iota_{K}\ast F - \Omega\wedge F .
\label{DiffSigma}
\end{eqnarray}
We now need to solve for the field strength $F$. In the process we shall see that the form of the metric is also fully determined by this set of equations.

If we begin by studying the consequences from the differential equation for $\Omega$~(\ref{DiffOmega}), we quickly find that, for example,

\begin{eqnarray}
\iota_{K}F_{01\alpha\beta} & = & \mp G(x^{i},z^{M},z^{\bar{N}},y) \left( dt \mp dx^1 \right) \wedge d\alpha \wedge d\beta \nonumber \\
d\Omega & = & \partial_{\chi} \left(f^2 g \right) \left( dt \mp dx^1 \right) \wedge dy \wedge d\chi .\nonumber
\end{eqnarray}
Setting $\alpha=y$ and $\beta=\chi$ we have

$$\mp G = \partial_{\chi} \left( f^2 g \right) \;\; \mbox{for} \; \chi = z^{M},z^{\bar{N}} .$$
This implies $G=0, \partial_{\chi} \left( f^2 g \right) = 0 \Longrightarrow f^2 g h(y) = constant$, where $h(y)$ is an arbitrary function of $y$. 

However, we can absorb this function into our metric co-ordinate $dy$ since that is the only place $g$ appears, so $gh(y)dy \rightarrow g dy^{\prime}$. Requiring that the metric is asymptotically Minkowski means we can set $f^2 g=1$ and therefore 
$$f^2=g^{-1}.$$ 

This reproduces, for example, the constraint $\partial_{\bar{N}} \ln g = -2 \partial_{\bar{N}} \ln f$ labelled as equation (8) in~\cite{Brinne:2000nf}. Both the components $F_{01y\beta}$ and its Hodge dual seven-form components $F_{23MNPQy}$ (where $MNPQ$ are a non-trivial combination of holomorphic and anti-holomorphic indices) then vanish. This also implies that the contraction

\begin{equation}
\iota_{\Omega}F = 0,
\end{equation}
which will simplify the calculations in what follows. We shall proceed in a similar manner in the analysis of the other differential equations.

The differential equation for $\Sigma$~(\ref{DiffSigma}) yields numerous results. Foremost among them are the non-trivial components of the field strength F

\begin{eqnarray}
F_{MN\bar{P}\bar{Q}} &=& \frac{i}{2}\partial_{y} \left[ f^2 \left( G_{N\bar{P}}G_{M\bar{Q}} - G_{N\bar{Q}}G_{M\bar{P}} \right) \right]\\
F_{PQ\bar{R}y} &=& \frac{i}{2} \left[ \partial_{P} \left( f^{-2} G_{Q\bar{R}} \right) - \partial_{Q} \left( f^{-2} G_{P\bar{Q}} \right) \right] 
\end{eqnarray}
and their complex conjugates. The second result is calculated from the Hodge dual seven form components $F_{0123M\bar{N}S}$, where we have used the conventions outlined in Appendix~\ref{Conventions}. 

There are also relations between the undetermined functions $f,g$ and the determinant of the hermitian part of the metric $G$. Concretely, defining a new function $H$, we have 

\begin{equation}
H^2 \vert P(z) \vert^2 \equiv g^{6} \vert P(z) \vert^2 = \vert \mbox{det G} \vert
\end{equation}
with $P$ an arbitrary holomorphic function of $z^{M}$. This allows for the freedom to make a holomorphic change of variables, in agreement with the observations of~\cite{Brinne:2000nf,Gomberoff:1999ps,Cho:2000hg}. In our co-ordinates we have chosen $P(z)=1$.

Furthermore, if we rescale the metric such that

\begin{equation}
g_{M\bar{N}} = H^{-1/6} G_{M\bar{N}}
\label{SmallgtoBigG}
\end{equation}
its associated hermitian 2-form becomes

\begin{equation}
\omega = ig_{M\bar{N}} dz^{M}\wedge dz^{\bar{N}}.
\end{equation}
In this form a further constraint derived from our differential equation can be succinctly written as in~(\ref{N1omwedgeomconstraint})

\begin{equation}
\partial \left( \bar{\partial} \right) \left( \omega\wedge\omega \right) = \partial \ast \omega = 0.
\label{N1Calbound}
\end{equation}
This co-K\"{a}hler calibration agrees with the constraints on generalised calibrations typical of these spacetimes~\cite{Husain:2003ag}. We shall use this constraint to calculate the K\"{a}hler metric on the moduli space of complex scalars of the $\mathcal{N}=1$ gauge theory in the next subsection.

One can check that the rest of the constraints listed as (6-13) in~\cite{Brinne:2000nf} are reproduced in their entirety. In addition, one must check that the equations of motion for the four-form field strength and the Bianchi identity are satisfied. 

Since we are considering an M5-brane geometry, which couples magnetically to the three-form potential, the roles of the Bianchi identity and the equation of motion are reversed. This means we require that $d\ast F = 0$ trivially. This is satisfied with the non-trivial components of $F$ we have calculated.

In summary, in terms of the rescaled metric, the solution is in agreement with~\cite{Brinne:2000nf} (as in~(\ref{N1Metric}), reproduced here for convenience):

\begin{eqnarray}
ds^2 &=& H^{-1/3} {dx^2}_{(1,3)} + 2 H^{1/6} g_{M \bar{N}} dz^{M}dz^{\bar{N}} + H^{2/3} dy^2  \\
F &=& \partial_{y}(\omega\wedge\omega)-i\partial(H^{1/2}\omega)\wedge dy + i\bar{\partial}(H^{1/2}\omega)\wedge dy .
\end{eqnarray}
The equation of motion for $F$ takes the form:

\begin{equation}
dF = \partial_{y} \left( \omega\wedge\omega\right)\wedge dy - 2i\bar{\partial}\partial \left( H^{1/2}\omega \right)\wedge dy = J,
\end{equation}
where $J$ denotes the source five-form specifying the shape of the Riemann surface describing our M5-brane configuration.

This method is quite similar in spirit to the G-structures approach developed in~\cite{Gauntlett:2002fz,Gauntlett:2003wb,Martelli:2003ki}, and it was shown in~\cite{Martelli:2003ki} how that group theoretic approach could rederive this same supergravity solution.

\subsection{Probe calculation of complex scalar moduli space}

We can perform a similar probe calculation to that of Section~\ref{ComScaKinTermsProbeN2} to determine the form of the moduli space of the complex scalars, although now in the $\mathcal{N}=1$ supersymmetric gauge theory case. The main difference is that now the M5-brane is probing a background of M5-branes wrapping 2-cycles in $\mathbf{C}^{3}$ (instead of $\mathbf{C}^{2}$). 

The metric we are using for this background, in terms of co-ordinates similar to the $F^2,G$ co-ordinates of the $\mathcal{N}=2$ case is

\begin{equation}
ds^2 = H^{-1/3} {dx^2}_{(1,3)} + 2 \left( H^{-1/3}\left\vert dG \right\vert^2 + H^{2/3}\left\vert dF^2 \right\vert^2 + H^{2/3}\left\vert dK^2 \right\vert^2 \right) + H^{2/3} dy^2 
\label{N1FGKmetric}
\end{equation}
where, as before, $F^2$ and $K^2$ can be thought of a local co-ordinates perpendicular to the background M5-brane, and $G$ as locally parallel to the background M5-brane. This implies that the Jacobian with respect to the $z^i$ ($i=1,2,3$) must be equal to one. It must also be the case that $H$ is harmonic in $F^2,K^2$ and $y$. One can check that this form of the metric satisfies the equations of motion. This could be used for explicit calculations but we will be able to show that $K_{\alpha\bar{\beta}}$ is K\"{a}hler from the general equations of motion~(\ref{N1Metric},\ref{N1omwedgeomconstraint}).

The holomorphic embeddings are now:

$$\begin{array}{rcl}
X^m & = & {\sigma}^{m}  \\
X^M & = & X^M \left( z, {\sigma}^{m} , u_{\alpha}\left({\sigma}^{m}\right)\right) \\
X^{\bar{N}} & = & X^{\bar{N}} \left( \bar{z}, {\sigma}^{m} , u_{\bar{\beta}}\left({\sigma}^{m}\right) \right) \\
X^{y} & = & X^{y} \left( z,\bar{z}, {\sigma}^{m} , u_{\alpha}\left({\sigma}^{m}\right) ,  u_{\bar{\beta}}\left({\sigma}^{m}\right) \right),
\end{array}$$
with $m=0\ldots 3$, $M,N=F^2,K^2,G$ and $y$ refers to $x^{(10)}$, the totally transverse direction. The $z,\bar{z}$ are arbitrary complex co-ordinates on the M5-brane worldvolume. The same arguments about small deviations from a supersymmetric embedding of our probe in the $X^y$ directions we used previously also apply in this case.

The calibration bound satisfied by our M5-brane probe is also different to the previous case where we probed a background of M5-branes wrapping a 2-cycle in $\mathbf{C}^{2}$. Then, it was a K\"{a}hler calibration which the probe had to satisfy. From the previous subsection, for a background of M5-branes wrapping a 2-cycle in $\mathbf{C}^{3}$, the calibration bound our probe has to satisfy is given by Equation~(\ref{N1Calbound}). In terms of the metric $G_{M\bar{N}}$ (which we recall is $G_{M\bar{N}}=H^{1/6}g_{M\bar{N}}$), the constraint takes the form

\begin{equation}
\partial_{[R} \left( H^{-1/3} \left( G_{M\vert\bar{P}\vert}G_{N\vert\bar{Q}\vert} - G_{M\vert\bar{Q}\vert}G_{N]\bar{P}} \right) \right) = 0.
\label{N1compcalbound}
\end{equation}
This constraint is an essential element in the calculation.

Repeating the analysis of Section~\ref{ComScaKinTermsProbeN2} reveals, with the appropriate extension to $\mathbf{C}^{3}$, the following action for the M5-brane probe

\begin{equation}
S = \int d^4\sigma d^2z \; H^{-2/3} g_{z\bar{z}} \sqrt{-\mbox{det}\left({\eta}_{mn} + H^{1/3} L_{mn}\right)}
\label{N1modspaceprobe}
\end{equation}
with $L_{mn}$ of the same form as before. Expanding this action and looking at the quadratic terms in the complex moduli only, we find

\begin{equation}
S_{kin} = \tau_{5} \int d^4\sigma \; {\partial}_{m} u_{\alpha} {\partial}^{m} u_{\bar{\beta}} K_{\alpha \bar{\beta}},
\label{KinComScalarsN1}
\end{equation}
where $K_{\alpha \bar{\beta}}$ is a K\"{a}hler metric given by

\begin{eqnarray}
K_{\alpha \bar{\beta}} &=& {\int}_{\Lambda} d^{2}z \; H^{-1/3} \left( G_{\alpha \bar{\beta}} G_{z \bar{z}} -  G_{\alpha \bar{z}}  G_{z \bar{\beta}} \right) \\
& =& {\int}_{\Lambda} d^{2}z \; T_{MN\bar{\beta}} \partial_{\alpha}X^{M} \partial_{z}X^{N}.
\label{KmetricComKinN1}
\end{eqnarray}
We have introduced the notation 

\begin{equation}
T_{MN\bar{\beta}} \equiv  H^{-1/3} \left( G_{M\bar{P}}G_{N\bar{Q}} - G_{M\bar{Q}}G_{N\bar{P}} \right) \left(\overline{\partial_{\beta}X^{P}}\right) \left(\overline{\partial_{z}X^Q}\right) .
\end{equation}
As before, $G_{\alpha \bar{\beta}} \equiv \frac{\partial X^M}{\partial u_{\alpha}} G_{M\bar{N}} \frac{\partial X^{\bar{N}}}{\partial u_{\bar{\beta}}}$, and $G_{M\bar{N}}$ refers to the spacetime metric~(\ref{SmallgtoBigG}). In our notation, we have that $\partial_{[P}T_{MN]\bar{\beta}} = 0$ (from~(\ref{N1Calbound}) or~(\ref{N1compcalbound})) and also that $T_{MN\bar{\beta}}=-T_{NM\bar{\beta}}$.

The form of the metric $K_{\alpha \bar{\beta}}$ is quite suggestive, and using the constraint~(\ref{N1compcalbound}), we can show this is K\"{a}hler up to total derivative boundary terms. Taking the anti-symmetrized derivative of this metric we get

\begin{eqnarray}
\partial_{[\gamma}K_{\alpha]\bar{\beta}} &=& {\int}_{\Lambda} d^{2}z \; \partial_{\gamma}\left(T_{MN\bar{\beta}}\right) \partial_{\alpha}X^{M} \partial_{z}X^{N} - {\int}_{\Lambda} d^{2}z \; \partial_{\alpha}\left(T_{MN\bar{\beta}}\right) \partial_{\gamma}X^{M} \partial_{z}X^{N} \nonumber \\
& &+ {\int}_{\Lambda} d^{2}z \; T_{MN\bar{\beta}} \partial_{\alpha}X^{M} \left(\partial_{\gamma}\partial_{z}X^{N}\right) + {\int}_{\Lambda} d^{2}z \; T_{MN\bar{\beta}} \partial_{\gamma}X^{N} \left(\partial_{\alpha}\partial_{z}X^{M}\right) \nonumber \\
& &
\end{eqnarray}
Integrating the third term by parts and simplifying the result we have

\begin{eqnarray}
\partial_{[\gamma}K_{\alpha]\bar{\beta}} &=& {\int}_{\Lambda} d^{2}z \; \left(\partial_{R}T_{MN\bar{\beta}}\right) \partial_{\alpha}X^{M} \partial_{z}X^{N} \partial_{\gamma}X^R \nonumber \\
& & - {\int}_{\Lambda} d^{2}z \; \left(\partial_{R}T_{MN\bar{\beta}}\right) \partial_{\gamma}X^{M} \partial_{z}X^{N}\partial_{\alpha}X^{R} \nonumber \nonumber \\
& &-{\int}_{\Lambda} d^{2}z \; \left(\partial_{R}T_{MN\bar{\beta}}\right) \partial_{\alpha}X^{M} \partial_{\gamma}X^{N}\partial_{z}X^R \nonumber \\
& &+ {\int}_{\Lambda} d^{2}z \; \partial_{z}\left(T_{MN\bar{\beta}} \partial_{\alpha}X^{M} \partial_{z}X^{N}\right) \nonumber \\
&=& \partial_{\alpha}X^{M} \partial_{z}X^{N}\partial_{\gamma}X^{R}\left[ \partial_{R}T_{MN\bar{\beta}}-\partial_{M}T_{RN\bar{\beta}}-\partial_{N}T_{MR\bar{\beta}}\right] \nonumber \\
& & + {\int}_{\Lambda} d^{2}z \; \partial_{z}\left(T_{MN\bar{\beta}} \partial_{\alpha}X^{M} \partial_{z}X^{N}\right) \nonumber \\
& = & {\int}_{\Lambda} d^{2}z \; \partial_{z}\left(T_{MN\bar{\beta}} \partial_{\alpha}X^{M} \partial_{z}X^{N}\right).
\end{eqnarray}
So again we have a total derivative for the boundary term, and the moduli space metric is indeed K\"{a}hler. An important role was played by the spacetime calibration bound~(\ref{N1compcalbound}) (or equivalently~(\ref{N1Calbound})) in analogy with the calculation of Section~\ref{ComScaKinTermsProbeN2}. However, note that in this case it is not a K\"{a}hler calibration but part of the more generalised calibrations typical of these spacetimes.

\subsection{The dimensional reduction of the background supergravity solution}

This follows very similar lines to the $\mathcal{N}=2$ case,  so again we just give the results. We find the eleven dimensional metric becomes

\begin{equation}
ds^{2}_{11} = g^{-1/3}dx^{2}_{(1,3)} + 2g^{1/6}M_{\mu\nu}dx^{\mu}dx^{\nu} + 2g^{1/6}ds^{2}_{KK} + g^{2/3}dy^{2}
\label{N111dKKmetric}
\end{equation}
where now

\begin{equation}
ds^{2}_{KK} = \frac{g_{7\bar{7}}}{\cos^{2}\theta}\left( d\tilde{x}_{7} + D_{\mu}dx^{\mu} \right)^{2} .
\end{equation}
If, as before, we denote the $\mu\nu=(k,l,\tilde{6})$, then the $M_{\mu\nu}$ part corresponds to 

\begin{eqnarray}
M_{\mu\nu}dx^{\mu}dx^{\nu} & = & \left( g_{k\bar{l}} - \frac{\cos^2\theta k_{k}k_{l}}{2g_{7\bar{7}}} \right) dx^{k}dx^{l} - \frac{\cos^2\theta k_{k}^2}{ 4 g_{7\bar{7}}} dx^{k}d\tilde{x}^{6} \nonumber \\
& & + \sin\theta k_{l}d\tilde{x}^6dx^{l} + \left( g_{\tilde{6}\tilde{\bar{6}}}  - g_{6\bar{6}}\tan^2\theta\right)d\tilde{x}^6d\tilde{x}^6
\label{N1MpartMetric}
\end{eqnarray}
where $k_{l}=\left[ (g_{l\bar{7}} + g_{7\bar{l}}) - \tan\theta(g_{l\bar{6}} + g_{6\bar{l}})\right]$. The difference is that we now have $k,l=4,5,8,9$.

Using the same form of the Kaluza-Klein reduction ans\"{a}tze~(\ref{KKmatrix}) used previously, we can easily read off the various supergravity fields. In particular, the dilaton is now

\begin{equation}
\phi = \frac{3}{4} \ln \left( \frac{2 g_{7\bar{7}} g^{1/6}}{ \cos^2\theta}\right)
\end{equation}
and the six component of the R-R one form becomes

\begin{equation}
C_{6}^{(1)} = -g^{-1/8}\sin\theta\left[\frac{8g_{7\bar{7}}^3}{\cos^6\theta}\right]^{(1/4)}.
\label{N110dOneForm}
\end{equation}
 
\subsection{The Yang-Mills coupling and theta angle}

The ten-dimensional string frame background metric is given by

\begin{equation}
ds^{2}_{10} = \frac{\sqrt{2g_{7\bar{7}}}}{\cos\theta}\left[H^{-1/4}dx^{2}_{(1,3)} + 2H^{1/4}M_{\mu\nu}dx^{\mu}dx^{\nu} + H^{3/4}dy^{2}\right] .
\end{equation}
If we now place our probe so that it lies along the $0123[\tilde{6}]$ directions, we can calculate the induced metric and thus the Yang-Mills coupling. This turns out to be 

\begin{equation}
\frac{1}{g^2_{YM}}= \frac{N}{8\pi^2 g_{s}l_{s}} \int d\tilde{x}^{6}  \;  \cos\theta
\label{N1YMcoupling}
\end{equation}
where to evaluate the determinant of the induced metric we have used $M_{\tilde{6}\tilde{6}}=g_{7\bar{7}}$, which follows from the hermitian condition of the metric components. Again we now include a factor of $N$ to take into account the possibility of the probe wrapping the $x^7$ direction $N$ times.

The theta angle turns out to be quite simple as well, explicitly

\begin{equation}
\Theta_{YM} = \left[ \frac{N}{g_{s}l_{s}} \int d\tilde{x}^{6}  \; (  \sin\theta) \right] \mod 2\pi N .
\label{N1YMthetaangle}
\end{equation}
We note that these results are in exact agreement with the previous $\mathcal{N}=2$ results. In the pure $\mathcal{N}=1$ Yang-Mills case, when all $N$ D4-branes are coincident due to the rotation of one of the NS5-branes, we again have a classical gauge coupling proportional to $L_{6}$, the separation of the NS5-branes at that particular point. The theta angle would be proportional to $L_{\theta}$, the distance between the endpoints of the D4-branes in the $x^7$ direction.

\subsection{Instantons}

We can also probe the $\mathcal{N}=1$ background along $\tilde{x}^6$ with Euclidean D0-brane probes to find the corresponding instanton action in the D4-brane world-volume gauge theory. This turns out to be exactly the same as for the $\mathcal{N}=2$ case, concretely

\begin{equation}
S_{D0} = \frac{8\pi^2}{g^2_{YM}} - i\Theta_{YM},
\end{equation}
which is the correct form of the instanton action.

\section{Discussion}
\label{Discussion}

In this paper we have used M-branes as probes of the supersymmetric 11-dimensional supergravity solutions~\cite{Fayyazuddin:1999zu,Brinne:2000fh,Brinne:2000nf} corresponding to M5-branes wrapping 2-cycles in $\mathbf{C}^{2}$ and $\mathbf{C}^{3}$. These probes have revealed interesting features about the corresponding $\mathcal{N}=2$ and $\mathcal{N}=1$ field theories. In general there were no unwanted supergravity corrections to field theory parameters such as the gauge coupling and theta angle from this analysis.

In the case of M5-brane probes, we have determined that the K\"{a}hler metric for the kinetic term of the complex scalars in the $\mathcal{N}=2$ effective Lagrangian receives no supergravity corrections. This is also true of the gauge coupling and theta angle parameters. The static M2-brane probe calculation, probing the BPS spectra and corresponding to a monopole in the field theory, also agrees with the usual calibration arguments. 

We demonstrated a new derivation of the supergravity solution~\cite{Brinne:2000nf} using a method where the projection conditions and spinor differential equations~(\ref{DiffK},\ref{DiffOmega},\ref{DiffSigma}) played a central role. We also analysed the $\mathcal{N}=1$ field theory related to M5-branes wrapping a 2-cycle in $\mathbf{C}^{3}$. All the results showed no supergravity corrections to the usual flat-space field theory analysis.

It would be interesting and useful to achieve a good understanding of the boundary conditions of finite probes with respect to the background M5-brane configuration and understand in what sense one can do such a calculation in M-theory, which in Type IIA corresponds to a finite D4-brane probe.

Other possibilities to try in the future include using non-static branes to probe further interesting features of the field theory. One example would be using M2-branes to probe the k-monopole moduli space of SU($N$) $\mathcal{N}=2$ gauge theories. There are also other phenomena of $\mathcal{N}=1$ MQCD that could be investigated such as domain walls and vortices.

\section{Acknowledgements}

JSL would like to thank Maria Dolores Loureda for hospitality and the Sir Richard Stapley Educational Trust Fund for partial support. Many thanks also to Nancy Cho for proofreading the manuscript.

\newpage
\appendix

\section{Conventions}
\label{Conventions}

In our conventions the epsilon tensor is defined such that

\begin{equation}
\epsilon_{\mu_{1}\ldots\mu_{n}}=1=g\epsilon^{\mu_{1}\ldots\mu_{n}} .
\end{equation}
We define the complex co-ordinates $z^{m}=x^{m}+iy^{m}$ and also define

\begin{equation}
\epsilon_{M_{1}\ldots M_{n}}\epsilon_{{\bar{N}}_{1}\ldots {\bar{N}}_{n}} dz^{M_{1}}\wedge \ldots  dz^{M_{n}} \wedge dz^{\bar{N}_{1}} \ldots \wedge dz^{\bar{N}_{n}} = \epsilon_{\mu_{1}\ldots\mu_{n}} dx^{1}\wedge dy^{1} \wedge \ldots \wedge dx^{n}\wedge dy^{n}.
\end{equation}
A differential form with holomorphic and anti-holomorphic indices is defined as

\begin{equation}
F = \frac{1}{p!q!} F_{M_{1}\ldots M_{p} {\bar{N}}_{1}\ldots {\bar{N}}_{q}} dz^{M_{1}}\wedge \ldots dz^{M_{p}} \wedge dz^{{\bar{N}}_{1}}\wedge \ldots dz^{{\bar{N}}_{q}} .
\end{equation}
The Hodge dual on a manifold of the form $\mathbf{R}^{(n)} \times \mathbf{Q}^{(n_{C})}$, with $\mathbf{R}^{(n)}$ being an n-dimensional Lorentzian manifold and $\mathbf{Q}^{(n_{C})}$ being a hermitian manifold of complex dimension $n_{C}$, is defined, for an $(r,p,q)$-form, as

\begin{eqnarray}
\ast F &=& \frac{\sqrt{g_{R}} \sqrt{G_{C}}}{r!m_{p}!n_{q}!(n-r)!(n_{C}-m_{p})!(n_{C}-n_{q})!} \epsilon^{\mu_{1}\ldots \mu_{r}}_{\mu_{r+1}\ldots \mu_{n}} \epsilon^{m_{1}\ldots m_{p}}_{{\bar{m}}_{p+1}\ldots {\bar{m}}_{n_{C}}}\epsilon^{{\bar{n}}_{1}\ldots {\bar{n}}_{q}}_{n_{q+1}\ldots n_{n_{C}}} \nonumber \\
& & dx^{\nu_{r+1}}\wedge \ldots \wedge dx^{\nu_{n}}  \wedge  dz^{n_{q+1}} \wedge \ldots \wedge dz^{n_{C}}\wedge dz^{{\bar{m}}_{p+1}} \wedge \ldots \wedge dz^{{\bar{m}}_{n_{C}}} .\nonumber \\
& & 
\end{eqnarray}
The determinant of a hermitian metric can be written in the form

\begin{eqnarray}
\sqrt{\det g_{M\bar{N}}} &=& \left( \frac{1}{n_{C}!}\right)^2 \epsilon^{M_{1}\ldots M_{n_{C}}} \epsilon^{\bar{N}_{1}\ldots \bar{M}_{n_{C}}} g_{M_{1}\bar{N}_{1}} \ldots g_{M_{n_{C}}\bar{N}_{n_{C}}} \\
&=& \left( \frac{1}{n_{C}!}\right)^2 \left\vert \epsilon^{M_{1}\ldots M_{n_{C}}} e^{1}_{M_{1}} \ldots e^{n_{C}}_{M_{n_{C}}} \right\vert^2 .
\end{eqnarray}
This implies that, for example, we have

\begin{equation}
e^{1}_{[1}e^{2}_{2}e^{3}_{3]} = 3! \left(\det g_{M\bar{N}}\right)^{1/4} .
\end{equation}
The component of the inverse metric compatible with this definition is

\begin{equation}
g^{M_{1}\bar{N}_{1}} = \frac{1}{(n_{C}-1)!} \frac{1}{\sqrt{\det g_{M\bar{N}}}} \epsilon^{M_{1}\ldots M_{n_{C}}} \epsilon^{\bar{N}_{1}\ldots \bar{N}_{n_{C}}} g_{M_{2}\bar{N}_{2}} \ldots g_{M_{n_{C}}\bar{N}_{n_{C}}} .
\end{equation}
A useful identity to know for the calculations of Section~\ref{AltSugraSol} is

\begin{equation}
\partial_{y} \left( f^{-4} g^{A\bar{B}} \right) = - \partial_{y} \left( f^{4} g_{M\bar{N}} \right) f^{-8} g^{M\bar{B}} g^{A\bar{N}} .
\end{equation}

\bibliographystyle{utphys}

\bibliography{references}

\providecommand{\href}[2]{#2}\begingroup\raggedright\begin{thebibliography}{10}

\bibitem{Fayyazuddin:1999zu}
A.~Fayyazuddin and D.~J. Smith, ``Localized intersections of M5-branes and
  four-dimensional superconformal field theories,'' JHEP {\bf 04} (1999) 030,
\href{http://xxx.lanl.gov/abs/hep-th/9902210}{{\tt hep-th/9902210}}.

\bibitem{Fayyazuddin:2000em}
A.~Fayyazuddin and D.~J. Smith, ``Warped AdS near-horizon geometry of
  completely localized intersections of M5-branes,'' JHEP {\bf 10} (2000) 023,
\href{http://xxx.lanl.gov/abs/hep-th/0006060}{{\tt hep-th/0006060}}.

\bibitem{Brinne:2000fh}
B.~Brinne, A.~Fayyazuddin, S.~Mukhopadhyay, and D.~J. Smith, ``Supergravity
  M5-branes wrapped on Riemann surfaces and their QFT duals,'' JHEP {\bf 12}
  (2000) 013,
\href{http://xxx.lanl.gov/abs/hep-th/0009047}{{\tt hep-th/0009047}}.

\bibitem{Brinne:2000nf}
B.~Brinne, A.~Fayyazuddin, T.~Z. Husain, and D.~J. Smith, ``N = 1 M5-brane
  geometries,'' JHEP {\bf 03} (2001) 052,
\href{http://xxx.lanl.gov/abs/hep-th/0012194}{{\tt hep-th/0012194}}.

\bibitem{Polchinski:1995mt}
J.~Polchinski, ``Dirichlet-Branes and Ramond-Ramond Charges,'' Phys. Rev. Lett.
  {\bf 75} (1995) 4724--4727,
\href{http://xxx.lanl.gov/abs/hep-th/9510017}{{\tt hep-th/9510017}}.

\bibitem{Polchinski:1995df}
J.~Polchinski and E.~Witten, ``Evidence for Heterotic - Type I String
  Duality,'' Nucl. Phys. {\bf B460} (1996) 525--540,
\href{http://xxx.lanl.gov/abs/hep-th/9510169}{{\tt hep-th/9510169}}.

\bibitem{Witten:1995ex}
E.~Witten, ``String theory dynamics in various dimensions,'' Nucl. Phys. {\bf
  B443} (1995) 85--126,
\href{http://xxx.lanl.gov/abs/hep-th/9503124}{{\tt hep-th/9503124}}.

\bibitem{Hull:1994ys}
C.~M. Hull and P.~K. Townsend, ``Unity of superstring dualities,'' Nucl. Phys.
  {\bf B438} (1995) 109--137,
\href{http://xxx.lanl.gov/abs/hep-th/9410167}{{\tt hep-th/9410167}}.

\bibitem{Townsend:1995kk}
P.~K. Townsend, ``The eleven-dimensional supermembrane revisited,'' Phys. Lett.
  {\bf B350} (1995) 184--187,
\href{http://xxx.lanl.gov/abs/hep-th/9501068}{{\tt hep-th/9501068}}.

\bibitem{Witten:1995im}
E.~Witten, ``Bound states of strings and p-branes,'' Nucl. Phys. {\bf B460}
  (1996) 335--350,
\href{http://xxx.lanl.gov/abs/hep-th/9510135}{{\tt hep-th/9510135}}.

\bibitem{Tseytlin:1996hi}
A.~A. Tseytlin, ``*No-force* condition and BPS combinations of p-branes in 11
  and 10 dimensions,'' Nucl. Phys. {\bf B487} (1997) 141--154,
\href{http://xxx.lanl.gov/abs/hep-th/9609212}{{\tt hep-th/9609212}}.

\bibitem{Gauntlett:1997cv}
J.~P. Gauntlett, ``Intersecting branes,''
\href{http://xxx.lanl.gov/abs/hep-th/9705011}{{\tt hep-th/9705011}}.

\bibitem{Ohta:1997fr}
N.~Ohta and P.~K. Townsend, ``Supersymmetry of M-branes at angles,'' Phys.
  Lett. {\bf B418} (1998) 77--84,
\href{http://xxx.lanl.gov/abs/hep-th/9710129}{{\tt hep-th/9710129}}.

\bibitem{Hanany:1996ie}
A.~Hanany and E.~Witten, ``Type IIB superstrings, BPS monopoles, and
  three-dimensional gauge dynamics,'' Nucl. Phys. {\bf B492} (1997) 152--190,
\href{http://xxx.lanl.gov/abs/hep-th/9611230}{{\tt hep-th/9611230}}.

\bibitem{Witten:1997sc}
E.~Witten, ``Solutions of four-dimensional field theories via M-theory,'' Nucl.
  Phys. {\bf B500} (1997) 3--42,
\href{http://xxx.lanl.gov/abs/hep-th/9703166}{{\tt hep-th/9703166}}.

\bibitem{Seiberg:1994rs}
N.~Seiberg and E.~Witten, ``Electric - magnetic duality, monopole condensation,
  and confinement in N=2 supersymmetric Yang-Mills theory,'' Nucl. Phys. {\bf
  B426} (1994) 19--52,
\href{http://xxx.lanl.gov/abs/hep-th/9407087}{{\tt hep-th/9407087}}.

\bibitem{Maldacena:1997re}
J.~M. Maldacena, ``The large N limit of superconformal field theories and
  supergravity,'' Adv. Theor. Math. Phys. {\bf 2} (1998) 231--252,
\href{http://xxx.lanl.gov/abs/hep-th/9711200}{{\tt hep-th/9711200}}.

\bibitem{'tHooft:1993gx}
G.~'t~Hooft, ``Dimensional reduction in quantum gravity,''
\href{http://xxx.lanl.gov/abs/gr-qc/9310026}{{\tt gr-qc/9310026}}.

\bibitem{Gubser:1998bc}
S.~S. Gubser, I.~R. Klebanov, and A.~M. Polyakov, ``Gauge theory correlators
  from non-critical string theory,'' Phys. Lett. {\bf B428} (1998) 105--114,
\href{http://xxx.lanl.gov/abs/hep-th/9802109}{{\tt hep-th/9802109}}.

\bibitem{Witten:1998qj}
E.~Witten, ``Anti-de Sitter space and holography,'' Adv. Theor. Math. Phys.
  {\bf 2} (1998) 253--291,
\href{http://xxx.lanl.gov/abs/hep-th/9802150}{{\tt hep-th/9802150}}.

\bibitem{Fayyazuddin:1997by}
A.~Fayyazuddin and M.~Spalinski, ``The Seiberg-Witten differential from
  M-theory,'' Nucl. Phys. {\bf B508} (1997) 219--228,
\href{http://xxx.lanl.gov/abs/hep-th/9706087}{{\tt hep-th/9706087}}.

\bibitem{Elitzur:1997fh}
S.~Elitzur, A.~Giveon, and D.~Kutasov, ``Branes and N = 1 duality in string
  theory,'' Phys. Lett. {\bf B400} (1997) 269--274,
\href{http://xxx.lanl.gov/abs/hep-th/9702014}{{\tt hep-th/9702014}}.

\bibitem{Elitzur:1997hc}
S.~Elitzur, A.~Giveon, D.~Kutasov, E.~Rabinovici, and A.~Schwimmer, ``Brane
  dynamics and N = 1 supersymmetric gauge theory,'' Nucl. Phys. {\bf B505}
  (1997) 202--250,
\href{http://xxx.lanl.gov/abs/hep-th/9704104}{{\tt hep-th/9704104}}.

\bibitem{Gauntlett:2002fz}
J.~P. Gauntlett and S.~Pakis, ``The geometry of D = 11 Killing spinors. ((T),''
  JHEP {\bf 04} (2003) 039,
\href{http://xxx.lanl.gov/abs/hep-th/0212008}{{\tt hep-th/0212008}}.

\bibitem{Gauntlett:2003wb}
J.~P. Gauntlett, J.~B. Gutowski, and S.~Pakis, ``The geometry of D = 11 null
  Killing spinors,'' JHEP {\bf 12} (2003) 049,
\href{http://xxx.lanl.gov/abs/hep-th/0311112}{{\tt hep-th/0311112}}.

\bibitem{Martelli:2003ki}
D.~Martelli and J.~Sparks, ``G-structures, fluxes and calibrations in
  M-theory,'' Phys. Rev. {\bf D68} (2003) 085014,
\href{http://xxx.lanl.gov/abs/hep-th/0306225}{{\tt hep-th/0306225}}.

\bibitem{Smith:2002wn}
D.~J. Smith, ``Intersecting brane solutions in string and M-theory,'' Class.
  Quant. Grav. {\bf 20} (2003) R233,
\href{http://xxx.lanl.gov/abs/hep-th/0210157}{{\tt hep-th/0210157}}.

\bibitem{Pasti:1997gx}
P.~Pasti, D.~P. Sorokin, and M.~Tonin, ``Covariant action for a D = 11
  five-brane with the chiral field,'' Phys. Lett. {\bf B398} (1997) 41--46,
\href{http://xxx.lanl.gov/abs/hep-th/9701037}{{\tt hep-th/9701037}}.

\bibitem{Aganagic:1997zq}
M.~Aganagic, J.~Park, C.~Popescu, and J.~H. Schwarz, ``World-volume action of
  the M-theory five-brane,'' Nucl. Phys. {\bf B496} (1997) 191--214,
\href{http://xxx.lanl.gov/abs/hep-th/9701166}{{\tt hep-th/9701166}}.

\bibitem{deBoer:1997zy}
J.~de~Boer, K.~Hori, H.~Ooguri, and Y.~Oz, ``Kaehler potential and higher
  derivative terms from M theory five-brane,'' Nucl. Phys. {\bf B518} (1998)
  173--211,
\href{http://xxx.lanl.gov/abs/hep-th/9711143}{{\tt hep-th/9711143}}.

\bibitem{Henningson:1997hy}
M.~Henningson and P.~Yi, ``Four-dimensional BPS-spectra via M-theory,'' Phys.
  Rev. {\bf D57} (1998) 1291--1298,
\href{http://xxx.lanl.gov/abs/hep-th/9707251}{{\tt hep-th/9707251}}.

\bibitem{Mikhailov:1997jv}
A.~Mikhailov, ``BPS states and minimal surfaces,'' Nucl. Phys. {\bf B533}
  (1998) 243--274,
\href{http://xxx.lanl.gov/abs/hep-th/9708068}{{\tt hep-th/9708068}}.

\bibitem{Harvey:1982xk}
R.~Harvey and J.~Lawson, H.~B., ``Calibrated geometries,'' Acta Math. {\bf 148}
  (1982)
47.

\bibitem{Gibbons:1998hm}
G.~W. Gibbons and G.~Papadopoulos, ``Calibrations and intersecting branes,''
  Commun. Math. Phys. {\bf 202} (1999) 593--619,
\href{http://xxx.lanl.gov/abs/hep-th/9803163}{{\tt hep-th/9803163}}.

\bibitem{Gauntlett:1998vk}
J.~P. Gauntlett, N.~D. Lambert, and P.~C. West, ``Branes and calibrated
  geometries,'' Commun. Math. Phys. {\bf 202} (1999) 571--592,
\href{http://xxx.lanl.gov/abs/hep-th/9803216}{{\tt hep-th/9803216}}.

\bibitem{Acharya:1998en}
B.~S. Acharya, J.~M. Figueroa-O'Farrill, and B.~Spence, ``Branes at angles and
  calibrated geometry,'' JHEP {\bf 04} (1998) 012,
\href{http://xxx.lanl.gov/abs/hep-th/9803260}{{\tt hep-th/9803260}}.

\bibitem{Hackett-Jones:2003vz}
E.~J. Hackett-Jones, D.~C. Page, and D.~J. Smith, ``Topological charges for
  branes in M-theory,'' JHEP {\bf 10} (2003) 005,
\href{http://xxx.lanl.gov/abs/hep-th/0306267}{{\tt hep-th/0306267}}.

\bibitem{Becker:1995kb}
K.~Becker, M.~Becker, and A.~Strominger, ``Five-branes, membranes and
  nonperturbative string theory,'' Nucl. Phys. {\bf B456} (1995) 130--152,
\href{http://xxx.lanl.gov/abs/hep-th/9507158}{{\tt hep-th/9507158}}.

\bibitem{Fayyazuddin:2005pm}
A.~Fayyazuddin, T.~Z. Husain, and I.~Pappa, ``The geometry of wrapped M5-branes
  in Calabi-Yau 2-folds,''
\href{http://xxx.lanl.gov/abs/hep-th/0509018}{{\tt hep-th/0509018}}.

\bibitem{DiVecchia:2002ks}
P.~Di~Vecchia, A.~Lerda, and P.~Merlatti, ``N = 1 and N = 2 super Yang-Mills
  theories from wrapped branes,'' Nucl. Phys. {\bf B646} (2002) 43--68,
\href{http://xxx.lanl.gov/abs/hep-th/0205204}{{\tt hep-th/0205204}}.

\bibitem{Aganagic:1997zk}
M.~Aganagic, J.~Park, C.~Popescu, and J.~H. Schwarz, ``Dual D-brane actions,''
  Nucl. Phys. {\bf B496} (1997) 215--230,
\href{http://xxx.lanl.gov/abs/hep-th/9702133}{{\tt hep-th/9702133}}.

\bibitem{Barbon:1997ak}
J.~L.~F. Barbon and A.~Pasquinucci, ``D0-branes, constrained instantons and D =
  4 super Yang- Mills theories,'' Nucl. Phys. {\bf B517} (1998) 125--141,
\href{http://xxx.lanl.gov/abs/hep-th/9708041}{{\tt hep-th/9708041}}.

\bibitem{Brodie:1998bv}
J.~H. Brodie, ``Fractional branes, confinement, and dynamically generated
  superpotentials,'' Nucl. Phys. {\bf B532} (1998) 137--152,
\href{http://xxx.lanl.gov/abs/hep-th/9803140}{{\tt hep-th/9803140}}.

\bibitem{Hori:1997ab}
K.~Hori, H.~Ooguri, and Y.~Oz, ``Strong coupling dynamics of four-dimensional N
  = 1 gauge theories from M theory fivebrane,'' Adv. Theor. Math. Phys. {\bf 1}
  (1998) 1--52,
\href{http://xxx.lanl.gov/abs/hep-th/9706082}{{\tt hep-th/9706082}}.

\bibitem{Witten:1997ep}
E.~Witten, ``Branes and the dynamics of {QCD},'' Nucl. Phys. {\bf B507} (1997)
  658--690,
\href{http://xxx.lanl.gov/abs/hep-th/9706109}{{\tt hep-th/9706109}}.

\bibitem{Giveon:1998sr}
A.~Giveon and D.~Kutasov, ``Brane dynamics and gauge theory,'' Rev. Mod. Phys.
  {\bf 71} (1999) 983--1084,
\href{http://xxx.lanl.gov/abs/hep-th/9802067}{{\tt hep-th/9802067}}.

\bibitem{Nunez:2003cf}
C.~Nunez, A.~Paredes, and A.~V. Ramallo, ``Flavoring the gravity dual of N = 1
  Yang-Mills with probes,'' JHEP {\bf 12} (2003) 024,
\href{http://xxx.lanl.gov/abs/hep-th/0311201}{{\tt hep-th/0311201}}.

\bibitem{Gomberoff:1999ps}
A.~Gomberoff, D.~Kastor, D.~Marolf, and J.~H. Traschen, ``Fully localized brane
  intersections: The plot thickens,'' Phys. Rev. {\bf D61} (2000) 024012,
\href{http://xxx.lanl.gov/abs/hep-th/9905094}{{\tt hep-th/9905094}}.

\bibitem{Cho:2000hg}
H.~Cho, M.~Emam, D.~Kastor, and J.~H. Traschen, ``Calibrations and
  Fayyazuddin-Smith spacetimes,'' Phys. Rev. {\bf D63} (2001) 064003,
\href{http://xxx.lanl.gov/abs/hep-th/0009062}{{\tt hep-th/0009062}}.

\bibitem{Husain:2003ag}
T.~Z. Husain, ``If I only had a brane!,''
\href{http://xxx.lanl.gov/abs/hep-th/0304143}{{\tt hep-th/0304143}}.

\end{thebibliography}\endgroup

\end{document}